\documentclass[floatfix,eqsecnum,nofootinbib,aps,12pt]{revtex4} 
\usepackage{epsfig}
\usepackage{subfigure}
\usepackage{graphicx}
\usepackage{hyperref}
\usepackage{color}

\def\eq#1{(\ref{#1})}

\def\s0#1#2{\mbox{\small{$ \frac{#1}{#2} $}}}
\def\0#1#2{\frac{#1}{#2}}

\usepackage{appendix}
\usepackage{amsmath}
\usepackage{amssymb}
\usepackage{graphicx}
\usepackage{amssymb}
\usepackage{epstopdf}
\usepackage{hyperref}
\usepackage{subfigure}
\usepackage{epstopdf}
\usepackage{verbatim} 
 \newcommand{\leftexp}[2]{{\vphantom{#2}}_{#1}{#2}}

\begin{document}

\title{Quantum gravity effects in Myers-Perry space-times}
\author{Daniel F. Litim}
\author{Konstantinos Nikolakopoulos}
\affiliation{Department of Physics and Astronomy, University of Sussex,\\ Brighton BN1 9QH, U.K.}
\date{\today}

\begin{abstract}\small We study quantum gravity effects for Myers-Perry black holes assuming that the leading contributions arise from the renormalization group evolution of Newton's coupling. Provided that gravity weakens following the asymptotic safety conjecture, we find that quantum effects lift a degeneracy of higher-dimensional black holes, and dominate over kinematical ones induced by rotation, particularly for small black hole mass, large angular momentum, and higher space-time dimensionality. Quantum-corrected space-times display inner and outer horizons, and show the existence of a black hole of smallest mass in any dimension. Ultra-spinning solutions no longer persist.
Thermodynamic properties including temperature, specific heat, the Komar integrals, and  aspects of black hole mechanics  are studied as well. Observing a softening of the ring singularity, we also discuss the validity of classical energy conditions.
\end{abstract}
\maketitle

\section{Introduction}

Black holes are classical solutions of the Einstein's field equations with many intriguing properties. Their most striking feature is the existence of a $2$-dimensional surface, the event horizon, which separates two causally disconnected regions of black-hole space-times. Since their first discovery by Schwarzschild \cite{Schwarzschild:1916uq}, further four-dimensional black hole solutions have been found including rotating Kerr black holes \cite{Kerr:1963ud}, or electrically charged ones.
Recent years have also seen an increased interest in the physics of higher-dimensional black holes, starting with 
the spherical symmetric solutions by Tangherlini \cite{Tangherlini} and the rotating ones by Myers and Perry \cite{MyersPerry}.
 The structure of higher-dimensional black holes appears to be much richer than their four-dimensional counterparts. Unlike the lower-dimensional cases, it was observed that rotating black holes in six and more space-time dimensions can have arbitrary large rotation. By now, it is understood that ultra-spinning solutions are classically unstable  \cite{instabus}, which has led to an extensive investigation of their instabilities \cite{Dias:2010eu,Dias:2009iu} and black hole phase transitions \cite{Gubser:2000mm,Dias:2010maa,Monteiro:2009tc,phasestructure}. Another remarkable feature of higher-dimensional solutions is the discovery of black objects with non-spherical event horizons \cite{blackrings}, which has as a direct consequence for the violation of uniqueness theorems in more than four space-time dimensions.

The description of black holes within general relativity is valid up to the semi-classical level, and expected to break down close to the Planck scale where quantum gravity effects become important. The perturbative non-renormalisability of gravity \cite{tHooft:1974bx} adds further obstacles to the study of quantum gravity effects. However, since the pioneering work of Weinberg \cite{Weinberg:1980gg} the possibility that gravity is non-perturbatively renormalisable has arisen and an increasing amount of evidence has been provided in four- \cite{Reuter:1996cp,Souma:1999at,Falls:2013bv} and  higher-dimensions \cite{Litim:2003vp,Litim:2006dx,Fischer:2006fz}, as well as in higher-derivative gravity \cite{Lauscher:2002sq,Codello:2007bd,Machado:2007ea,Benedetti:2009rx} and under the inclusion of matter fields \cite{Percacci:2002ie,Folkerts:2011jz,Harst:2011zx,Zanusso:2009bs}; for overviews see \cite{Litim:2011cp,Litim:2008tt,Fischer:2006at,Niedermaier:2006wt,Codello:2008vh}. Furthermore, a  bootstrap search strategy is available allowing self-consistent tests of the asymptotic safety conjecture \cite{Falls:2013bv}. Asymptotic safety offers a UV completion for metric quantum gravity due to a non-trivial high energy fixed point under the renormalisation group. 
By now, phenomenological studies of asymptotic safety for black hole physics have covered four- \cite{reuter} and higher-dimensional Schwarzschild black holes \cite{kevin}, Kerr black holes \cite{Reuter:2006rg,tuiran}, black holes from higher-derivative gravity \cite{Cai:2010zh}, the inclusion of boundary terms \cite{Becker:2012js}, quantum-corrected laws of black hole thermodynamics \cite{kevin2}, and black holes with a cosmological constant \cite{Bennett,Koch:2013owa}.

In this paper we investigate quantum corrections for rotating black holes in higher dimensions. Our primary interest relates to the interplay between kinematical effects induced by rotation, and quantum gravity effects induced by the energy-dependence of Newton's coupling.  Specifically, we wish to understand how the  `phase space'  of black holes ---  the set of black hole masses, angular momenta, and space-time dimensionality $(M,J,d)$ for which a black hole exists ---  is modified in the presence of quantum gravity effects. 
Our working hypothesis is that the main quantum effects  arise from the substituting the gravitational coupling with its running, scale-dependent counterpart as provided by the renormalisation group equations for gravity. For most parts, the energy-dependence of Newton's coupling will be imported from the asymptotic safety programme, where the short-distance behaviour is dictated by a gravitational fixed point $g_*$. The quantum-corrected `phase-space' is then characterised by $(M,J,d; g_*)$, and allows a smooth interpolation between the well-known classical results in the limit $g_*\to \infty$, and asymptotically safe black holes where $g_*$ is of order unity.  We also discuss, whenever possible,  implications arising from generic renormalisation group equations for gravity outside the asymptotic safety scenario.
 
Interestingly, our modifications can also be understood as arising from an effective energy momentum tensor,  whose properties are analysed in the light of the classical energy conditions and curvature singularities. Thereby we discuss in detail the RG-improved horizon, temperature, specific heat, as well as the singularity structure and modifications of the laws of black hole mechanics.

We organise the paper as follows. In Sec.~\ref{Generalities} we review classical higher-dimensional spinning black holes and introduce our setup for the inclusion of quantum corrections. In Sec.~\ref{Horizons} we investigate how the horizon structure is modified in four, five, and more space-time dimensions. In Sec.~\ref{Thermodynamics} we are concerned with aspects of black hole thermodynamics and examine the form of temperature and specific heat. In Sec.~\ref{Mass and energy} we derive and analyse the effective energy-momentum tensor and compute the Komar mass and angular momentum of the space-time. In Sec.~\ref{Conclusions} we summarize and discuss our findings. Three appendices contain technical aspects and extra material regarding the horizon structure (App.~\ref{Appendix}), the effective energy-momentum tensor (App.~\ref{Appendix2}), and the Kretschmann invariant (App.~\ref{singularities}).

\section{Generalities}\label{Generalities} 

In this section we review the classical higher dimensional rotating black hole solutions and we establish our notation. Furthermore, we present some considerations about quantum corrections and we define the setup we will use in this paper.

\subsection{Myers-Perry black holes}\label{review}
The first extension of solutions to Einstein's equations in higher dimension was made by Tangherlini in 1963, who generalised the spherically symmetric solutions, leading to the Schwarzschild-Tangherlini metric \cite{Tangherlini}. It was not until 1986 that spinning black holes in higher dimensions were found by Myers and Perry  \cite{MyersPerry}. Unlike the four dimensional case, spinning black holes in higher dimensions can rotate in more than one independent planes. For a complete review see \cite{Emparan}. For simplicity, here we are going to examine only the case of rotation in a single plane. This space-time is described by the metric
\begin{equation}
\begin{split}
ds^2 =&  -dt^2 +G_N\frac{M}{r^{d-5}\Sigma}(dt-a \sin^2\theta \,d\phi)^2+\frac{\Sigma}{\Delta}dr^2\\
&+\Sigma\, d\theta^2  +(r^2+a^2)\sin^2\theta\, d\phi^2  +r^2\cos^2\theta\, d\Omega^2_{d-4},
 \end{split}\label{MP}
\end{equation}\\
where $\Sigma$ and $\Delta$ are defined by
\begin{equation}
\begin{split}
\Sigma&=r^2+a^2\cos^2\theta\\ 
\frac{\Delta}{r^2}&=1+\frac{a^2}{r^2}-G_N\frac{M}{r^{d-3}}
\label{D}
\end{split}
\end{equation}
and $d$ is the number of space-time dimensions, $G_N$ the Newton's coupling constant, $d\Omega^2_{d-4}$ the line element on the unit $d-4$ sphere, while the reduced mass $M$ and the parameter $a$ are related to the physical mass $M_{\textnormal{phys}}$  and angular momentum $J$  by
\begin{eqnarray}
M&=&\frac{8\,\Gamma(\frac12(d-1))}{(d-2)\pi^{(d-3)/2}}\,M_{\textnormal{phys}}\\
 a&=&\frac{d-2}{2}\frac{J}{M_{\textnormal{phys}}}\,.
 \end{eqnarray}
The limit $d=4$ of this space-time is the well known Kerr solution. 
The horizons of Myers-Perry solutions are found from the coordinate singularities $g^{rr}=0$, or $\Delta=0$. 
We  observe  from \eqref{D} that $d=5$ is distinguished: 
The centrifugal force does not depend on the dimensionality of space-time.
On the other hand, the gravitational force is dimension-dependent and dominates for $d>5$ for small $r$, making \eqref{D} negative. This leads to event horizons for space-times with arbitrary large angular momentum, the so-called \emph{ultra-spinning} black holes. It is interesting to note that ultra-spinning regions can also exist when we have rotation in many planes \cite{Emparan}, \cite{MyersPerry}.

The above considerations make it interesting to study separately the properties of the three qualitatively different cases of $d=4$, $d=5$ and $d\geq6$. We begin with the four dimensional case where the line element \eqref{MP} reduces to that of the Kerr solution. The expression $\Delta(r)$ which gives the horizons is a quadratic polynomial with two solutions
\begin{equation}
r_{\pm}=\frac12\left[G_NM\pm \sqrt{(G_NM)^2-4a^2}\right],\label{Kerrsol}
\end{equation}
where $r_+$ corresponds to the event horizon and $r_-$ to the inner (Cauchy) horizon. It is evident from \eqref{Kerrsol} that horizons exist for black hole masses large enough to satisfy $G_NM\geq2a$. We observe that in four dimensions the effect of angular momentum has two important consequences to the structure of black holes. Firstly, black hole solutions exist for masses $M$ up to a kinematically-induced Kerr mass $M_{\textnormal{Kerr}}(a)\ge 0$.  Secondly, there exists an inner horizon.

In the five dimensional case, the term which is responsible for the gravitational attraction becomes constant and $\Delta(r)$ has only one positive root given by
\begin{equation}
r_+=\sqrt{G_NM-a^2}\,.
\label{5dsol}
\end{equation}
Hence, black hole solutions exist only for sufficiently large masses $G_N\,M>a^2$ and the resulting space-time has only one horizon.

In six or higher dimensions none of the two main features  of the $4d$ Kerr black holes are preserved: The gravitational term of $\Delta(r)$ dominates as we approach $r\to0$ and thus $\Delta(r)$ is always negative in this limit. This, together with the limit of $\Delta(r)$ when $r\to\infty$, always implies the existence of at least one horizon for any mass, with no further constraints due to angular momentum. 
Moreover, a check on the first derivative of $\Delta(r)$ with respect to $r$ reveals exactly one event horizon, and no inner horizon.

A generic feature of rotating black hole solutions is that, for given mass, their horizon radii are always less than their non-rotating counterparts. 
In fact, the horizon radius of the non-rotating Schwarzschild-Tangherlini black hole is given by
\begin{equation}
r_{\textnormal{cl}}=(G_NM)^{\frac{1}{d-3}}.\label{sw-tang}
\end{equation} 
Substituting \eq{sw-tang} into \eqref{D} gives $\Delta(r_{cl})=a^2\geq0$. For $d\geq5$, since the first derivative $\Delta'(r)$ is always positive, it follows that the horizons of spinning black holes at given mass $M$ (if they exist) are smaller of their non-rotating counterparts, $r_{\rm cl}(a)<r_{\rm cl}$. For $d=4$, this observation can be read off  from the horizons of the Schwarzschild and Kerr solutions \eq{Kerrsol},  
$r_+\leq G_NM=r_{\textnormal{cl}}$.

\begin{center}
\begin{table*}[t]
\begin{tabular}{c|c|c|c|c} 
${}\quad$case${}\quad$ 
&${}\quad$ 
index${}\quad$ 
& gravity 
& horizons & ${}\quad$ $\Delta(r\to0)$${}\quad$
\\
 \hline 
(i) & $s<d-5$ 
& ${}{}\quad $strong if $s<0$; 
weak if $s>0\quad$ 
& one or more & singular
\\[.5ex]
(ii) & $s=d-5$ 
& ${}{}\quad $strong if $s<0$; 
weak if $s>0\quad$ 
& none, one or more & finite
\\[.5ex]
(iii) & $s>d-5$ 
& ${}{}\quad $strong if $s<0$; 
weak if $s>0\quad$ 
& ${}\quad$ none, one or more $\quad$&  $a^2$ 
\end{tabular}  
\caption{Horizons of rotating black holes assuming a scale-dependent gravitational coupling strength 
\eq{char} at short distances for various dimensions and in dependence on a phenomenological short distance index $s$
 (see text).
\label{strongweakgravity}} 
\end{table*} 
\end{center}

\subsection{Quantum effects}\label{quantum corrections}

We expect that this picture changes when quantum gravity effects are taken into account and the black hole mass approaches the fundamental scale of quantum gravity. In what follows we will assume that the leading order quantum effects are captured from the renormalization of the gravitational coupling, which makes it a function of the momentum scale. We are going to  study these effects mostly in the context of the asymptotic safety scenario. 
Prior to this, we would like to gain a qualitatively picture.
Depending on the ultraviolet completion of gravity and the renormalization group, gravity may become "weaker" or "stronger" at short distances, or may be superseeded by an altogether different description. To investigate the impact on black holes for these cases, following \cite{kevin}, we  parametrize the putative running of Newton's constant at horizon scales as
\begin{equation}\label{char}
G_N\to G(r)=r^{d-2}_{\textnormal{char}}\left(\frac{r}{r_{\textnormal{char}}}\right)^s\,.
\end{equation}
Here, $r_{\textnormal{char}}$ denotes the characteristic length scale of the problem, $e.g.$ the horizon if it exists, and the phenomenological index $s$ parametrizes whether the gravitational coupling remains classical $(s=0)$, decreases ($s>0$), or increases ($s<0$) due to fluctuations. Then we substitute this form of $G(r)$ back to the expression \eqref{D} for $\Delta(r)$, 
and  distinguish the following three cases 
\begin{enumerate}
\item[(i)] $s<d-5$. The strength of the gravitational coupling  increases (decreases) for negative (positive) $d-5$.
The function $\Delta(r\to 0)$ is unbounded from below implying the existence of, at least, one horizon. 
\item[(ii)] $s=d-5$. In this case, we find a finite limit $\Delta(r\to 0)\equiv\Delta_0= a^2-r_{\textnormal{char}}^{3}M$. For $\Delta_0<0$, this necessarily implies the existence of a horizon, similar to case (i). For $\Delta_0>0$, the situation is similar to case (iii).
\item[(iii)] $s>d-5$. In this case the gravitational coupling becomes weaker for all $d\geq5$. For $d<5$, gravity may even become strong. In either case  $\Delta_0=a^2>0$ implying that the space-time can have none, one or more horizons, depending on the precise short distance behavior of $G(r)$ and the other parameters of the space-time such as the mass and angular momentum. 
\end{enumerate}
The conclusions of the above considerations are summarized in Tab.~\ref{strongweakgravity}

\subsection{Asymptotically safe gravity}\label{Asymptotic safety}

We now turn to a specific quantum gravity scenario known as asymptotic safety. The asymptotic safety conjecture was first proposed by Steven Weinberg \cite{Weinberg:1980gg} (for reviews, see \cite{Litim:2006dx,Litim:2008tt,Litim:2011cp}) and stipulates the existence of a non-trivial UV fixed point for gravity which governs the renormalisation group flow of gravity. The conjecture was shown to hold true perturbatively in the vicinity of two dimensions \cite{Weinberg:1980gg}, with subsequent studies extending it non-perturbatively to four \cite{Souma:1999at,Reuter:1996cp,Litim:2003vp} and higher dimensions \cite{Litim:2003vp,Fischer:2006fz,Litim:2006dx}. Furthermore, the appicability of a bootstrap search strategy for the fixed point has been established in \cite{Falls:2013bv} showing the exisatence of a fixed point for polynomial actions of the Ricci scalar $R$ up to the order $R^{34}$. It is thus conceivable that the fixed point exists in the full physical theory. In addition, the theory displays a Gaussian (non-interacting) fixed point and admits  RG trajectories which connect the UV and IR fixed points so that classical general relativity and conventional perturbation theory are recovered in the low-energy (long-distance) limit. 

For our present purposes it is sufficent to consider the non-perturbative RG flow of Newton's coupling asuming that the underlying action is given by the $d-$dimensional Ricci scalar \cite{Litim:2003vp,Litim:2001up}. 
The RG flow takes the form
\begin{equation}\label{eta}
k\partial_k g=(d-2+\eta)g
\end{equation}
where the dimensionless coupling $g=G_k\,k^{d-2}$ denotes the running Newton coupling in units of the RG scale $k$, and $\eta$ is the graviton anomalous dimension $\eta = -k\partial_k (G_k/G_N)$. The RG flow displays two types of fixed points, a Gaussian infrared fixed point at $g_*=0$, where gravity becomes classical  and $\eta=0$, and a non-Gaussian ultraviolet fixed point where $g_*\neq0$ and $\eta=2-d$. On RG trajectories connecting these fixed points, we note that the anomalous dimension interpolates between $0$ in the IR and $2-d$ in the UV. Upon integration of the higher-dimensional renormalisation group flow we find an implicit equation for the scale-dependent gravitational coupling given by \cite{Litim:2003vp,Litim:2001up}
\begin{equation}\label{Gk}
\frac{G_k}{G_N}=\left(1-\frac{G_k}{g_*\,k^{2-d}}\right)^\delta
\end{equation}
where 
$\delta=\theta_{\rm G}/\theta_{\rm NG}$ is the ratio of the universal scaling exponents $\theta$ at the Gaussian (non-Gaussian) fixed point, respectively and $g_*$ is the value of the UV fixed point. In \cite{Litim:2003vp,Fischer:2006fz}, it was found that $\theta_{\rm G}=d-2$ and $\theta_{\rm NG}=2d\,\frac{d-2}{d+2}$, leading to
\begin{equation}\label{delta}
\delta=\frac{2d}{d+2}\,.
\end{equation}
Hence, the index $\delta$ interpolates between 1 and 2 for $d\in[4,\infty]$. In the linear approximation $\theta_{\rm G}=\theta_{\rm NG}$ we have $\delta=1$, and consequently $G_k=G_N/(1+k^{d-2}/g_*)$. In the quadratic approximation, $\theta_{\rm G}=\frac12 \theta_{\rm NG}$  \cite{Gerwick:2011jw}. To conclude, for what concerns us here, the quantum gravity effects are parametrised by $g_*$, typically of order unity, and reduce to classical physics in the limit $g_*\to\infty$.

The input from asymptotic safety provides us with the running Newton coupling as a function of the RG momentum scale. However, black hole solutions are found in coordinates of curved space-time and we need to use a matching between momentum and position scales in order to make a replacement of Newton's constant, by a distance-dependent coupling
\begin{equation}
G_N\to G(r) \label{replacement}
\end{equation}
It is the central assumption this paper that the leading quantum gravity corrections originate from this replacement. Following \cite{kevin}, we will identify $k=\xi/r$, where $\xi$ is a non-universal parameter which also depends on the specific RG scheme used in the derivation of \eq{Gk} (see \cite{kevin2} for alternative matching conditions). Below, we mostly use the linear approximation \eq{Gk}, \eq{delta} with $\delta=1$. In this case the running of gravitational coupling \eqref{Gk} takes the form
\begin{equation}
G(r)=\frac{G_Nr^{d-2}}{r^{d-2}+{G_N/g_*}}\label{Gofr}
\end{equation}
where we tacitly assume that the matching parameter $\xi$ has been absorbed into $g_*$.

\section{Horizons}\label{Horizons}

After these preliminaries we study the horizon structure of black holes using a running for the Newton's constant dictated by the asymptotic safety conjecture. We recall that classical Myers-Perry black holes display two horizons under an auxiliary condition for $d=4$, one horizon under an auxiliary condition for $d=5$ and always a single horizon when $d\geq 6$. This pattern will be modified under quantum corrections, and the existence of horizons will depend on the precise form of $G(r)$. We start this section by presenting an analysis about the existence of horizons and their conditions. We are also interested to see if the possibility of ultra-spinning solutions will persist. Then, following from the fact that in the classical case $d=5$ is a critical dimensionality, we examine separately the cases $d=4$, $d=5$ and $d\geq6$.

\subsection{Horizon structure}\label{generalremarks}

First, we examine how many horizons arise in the presence of quantum corrections. For this, we recall the structure of non-rotating black holes within asymptotic safety and we investigate the modifications due to rotation, while keeping $G(r)$ as general as possible.  We concentrate on the horizon condition arising through  $\Delta(r)=0$. It is convenient to study the  roots of the dimensionless function 
\begin{equation}
f(r)=\frac{\Delta(r)}{r^2}\equiv1+\frac{a^2}{r^2}-\frac{M\,G(r)}{r^{d-3}}. \label{qualquant}
\end{equation}
In contrast to the classical case \eqref{D}, the running of gravitational coupling $G(r)$ weakens  the gravitational potential  in the context of asymptotic safety  at short distances. For given mass $M$ and angular momentum $a$, the RG improved horizon radius $r_s(M,a)$ is obtained as the implicit solution(s) of
\begin{equation}
r_s^{d-3}(M,a)=M\,G(r_s(M,a))-a^2\,r_s^{d-5}(M,a)\,,
\end{equation}
provided it exists.
\begin{figure*} [t]
\centering
\includegraphics[width=.49\hsize]{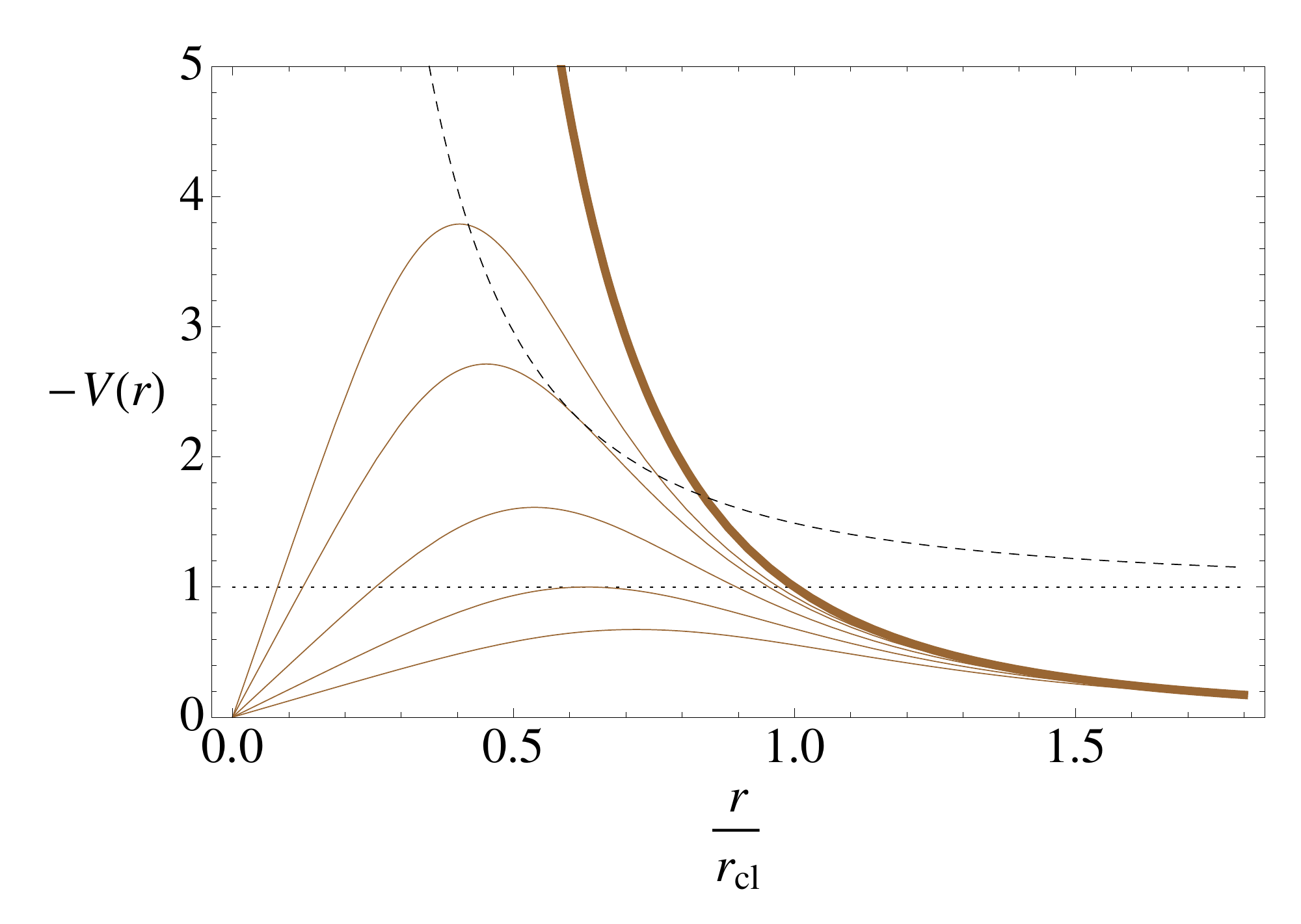}\hskip.01\hsize
\includegraphics[width=.49\hsize]{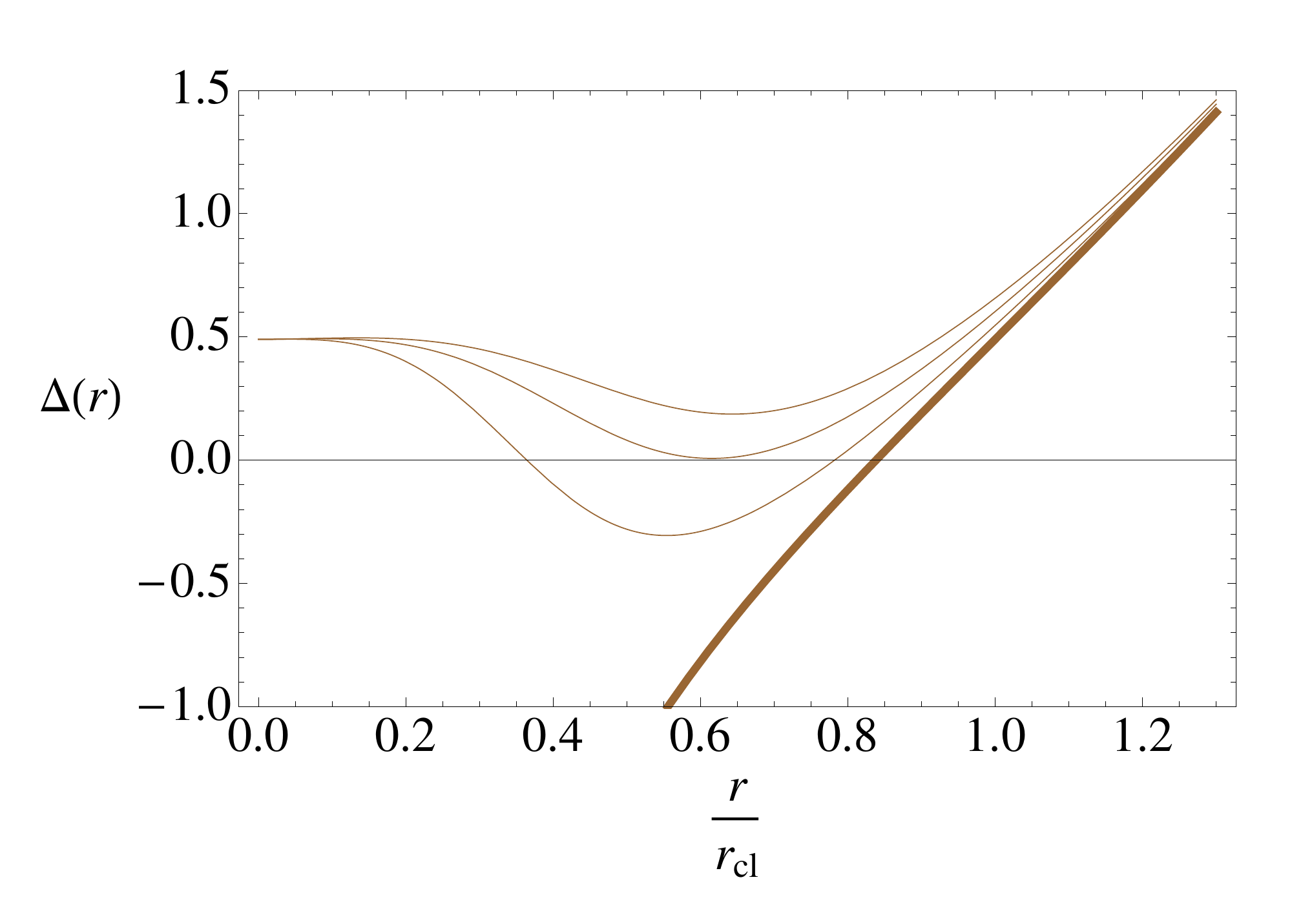}\caption{\label{Hor1} Left panel: The gravitational potential $-MG(r)/r^{d-3}$ (full lines) in comparison with the rotational barrier $1+{a^2}/{r^2}$ ($a=0:$ dotted line, $a\neq 0$: dashed line). Right panel: the function $\Delta(r)$ \eq{qualquant} for a theory where gravity becomes weaker at small distances ($d=6$ and $a=0.7\, M_{\textnormal{Pl}}^{-2}$). In either case, the thick (brown) line denotes the classical limit, and thin lines denote decreasing values for $M$  (top to bottom) in the case where $G(r)$ is weakening towards shorter distances.}
\end{figure*} 
Assuming that $M\,G(r)\,r^{3-d}$ vanishes in the limit $r\to0$, we can deduce the form of the gravitational potential $V(r)=-M\,G(r)\,r^{3-d}$. Its first derivative with respect to $r$ is given by
\begin{equation}
V'(r)=M\, G(r)\, r^{2-d}\left( d-3+\eta(r)\right),\label{fnonrot}
\end{equation}
where $\eta(r)=-r\, G'(r)/G(r)$ is the anomalous dimension of the graviton after scale identification. As implied by the previous section, in the context of asymptotic safety, $\eta(r)$ is a monotonically increasing function, which interpolates from $\eta(0)=2-d$ to $\eta(\infty)=0$. Thus, it is obvious from \eqref{fnonrot} that $V'(r)$ changes sign once and $V(r)$ decreases from $V(0)=0$ down to a minimum value $V_{\textnormal{min}}$ and then it increases back to $V(\infty)=0$. This behaviour reflects the weakening of gravity in our model, see Fig. \ref{Hor1}.
In the absence of rotation, $V(r)$ competes the constant barrier $1$. Thus, if $V_{\textnormal{min}}<-1$ the space-time has two horizons, if $V_{\textnormal{min}}=-1$ it has one degenerate horizon, while if $V_{\textnormal{min}}>-1$ there are no horizons. Which of the three cases is actually realised depends on the mass $M$ and the precise form of $G(r)$ \cite{reuter,kevin}.
When we consider rotating black holes, the gravitational potential $V(r)$ competes the constant term enhanced by the rotational term, see Fig. \ref{Hor1}. In order to examine how many horizons we have, we look for roots of the derivative of $f(r)$, given by
\begin{equation}
f'(r)=r^{-3}\left[-2a^2+M\, G(r)\, r^{5-d}\left( d-3+\eta(r)\right)\right]. \label{qualquantp}
\end{equation}
Using the same assumptions as for the non-rotating case, we find that in four and five dimensions $f(r)$ has only one minimum and the space-time can have either two, one degenerate, or no horizons. For six or higher dimensions we have to know more details about $G(r)$, but for the running of Newton's coupling given by \eqref{Gofr}, the same behaviour with either two, one degenerate or no horizons still holds (for details see Appendix \ref{Appendix}). The resulting form of $\Delta(r)$ is plotted in Fig. \ref{Hor1}.

\subsection{Critical mass}
This behaviour of the function $f(r)$, implies that black hole solutions exist only for masses greater than a minimum mass $M_{\textnormal{c}}$. For non-rotating black holes within asymptotic safety there is a map between the RG parameter $g_*$ and a black hole which is characterised by some $M_{\textnormal{c}}$. As we shall see in more detail later, for rotating black holes, $M_{\textnormal{c}}$ is also a function of angular momentum and so there is a map between the parameter $g_*$ in \eq{Gofr} and the smallest achievable black hole mass $M_{\textnormal{c}}(a)$ in dependence of angular momentum. Black holes exist only for
\begin{equation}
M\geq M_{\textnormal{c}}(a).
\end{equation} 
For the rest of this paper we will write $M_{\textnormal{c}}$ for the critical mass of the non-rotating black hole \cite{kevin}, while for a rotating black hole we will write explicitly $M_{\textnormal{c}}(a)$.
Moreover, when we encounter the RG parameter $g_*$, we will eliminate it in favour of $M_{\textnormal{c}}(a)$. This choice allows us to compare results with other models of quantum gravity exhibiting a smallest black hole mass $M_{\textnormal{c}}$ without necessarily arising through an underlying fixed point $g_*$ (see for example \cite{Nicolini:2005vd}, \cite{Modesto:2010rv}).

Keeping the RG function $G(r)$ arbitrary for the time being, we solve simultaneously the conditions $f(r)=0$ and $f'(r)=0$ using \eqref{qualquant} and \eqref{qualquantp} to find the value of the anomalous dimension for an RG improved black hole with a critical horizon,
\begin{equation}
\eta(r_{\textnormal{c}})=3-d+\frac{2a^2}{r_{\textnormal{c}}^2+a^2}\,.\label{EtaCritical}
\end{equation}
Here $r_{\textnormal{c}}$ denotes the radius of the critical (degenerate) horizon. This result is to be compared with the non-rotating case where $\eta(r_{\textnormal{c}})=3-d$ \cite{kevin}. Thus, without relying on any specific form for the running of Newton's coupling, we conclude that the graviton anomalous dimension for a critical rotating black hole will be smaller in magnitude than the one of the corresponding spherical black hole. In general it will satisfy
\begin{equation}
3-d\leq \eta(r_{\textnormal{c}}) < 5-d\,.
\end{equation}
Next, we will analyse this more quantitatively using suitable dimensionless parameters.

\subsection{Critical parameters}

In order to quantitatively examine the properties of RG corrected black holes we express the relevant equations in terms of dimensionless variables. This is achieved by dividing every dimensionfull parameter by the appropriate power of a representative length, which we take to be the horizon radius of classical, non-rotating black holes, i.e. $r_{\textnormal{cl}}=(G_NM)^{\frac{1}{d-3}}$.  The mass dimensions of the parameters are $[G_N]=2-d$, $[a]=-1$ and $[M]=1$, leading to
\begin{equation}\label{x}
x=\frac{r}{r_{\textnormal{cl}}}\,.
\end{equation}
The parameter $x$ measures the radial coordinate in units of the horizon of {\em classical non-rotating} black holes $r_{\textnormal{cl}}$ correspondong to the same mass $M$. If they exist, we denote the event horizon by $x_+$, the Cauchy (inner) horizon by $x_-$, and the degenerate (critical) horizon by $x_c$. 
The parameter 
\begin{equation}\label{A}
A=\frac{a^2}{r_{\textnormal{cl}}^{2}} 
\end{equation}
provides the dimensionless ratio of the angular momentum parameter $a$ in units of the classical Schwarzschild horizon. 
Finally, we introduce the parameter
\begin{equation}\label{dimlesspar}
\Omega
=\frac1{g_*\,M\,r_{\rm cl}(M)}\,.
\end{equation}
It measures the deviation from classical gravity to which our equations reduce for $\Omega\to 0$. Note that the classical limit is achieved by sending the black hole mass to infinity $M/M_P \to\infty$ for fixed $g_*$, or by sending $g_*\to \infty$ for fixed mass. 

\begin{figure} [t]
\centering
\includegraphics[width=.6\hsize]{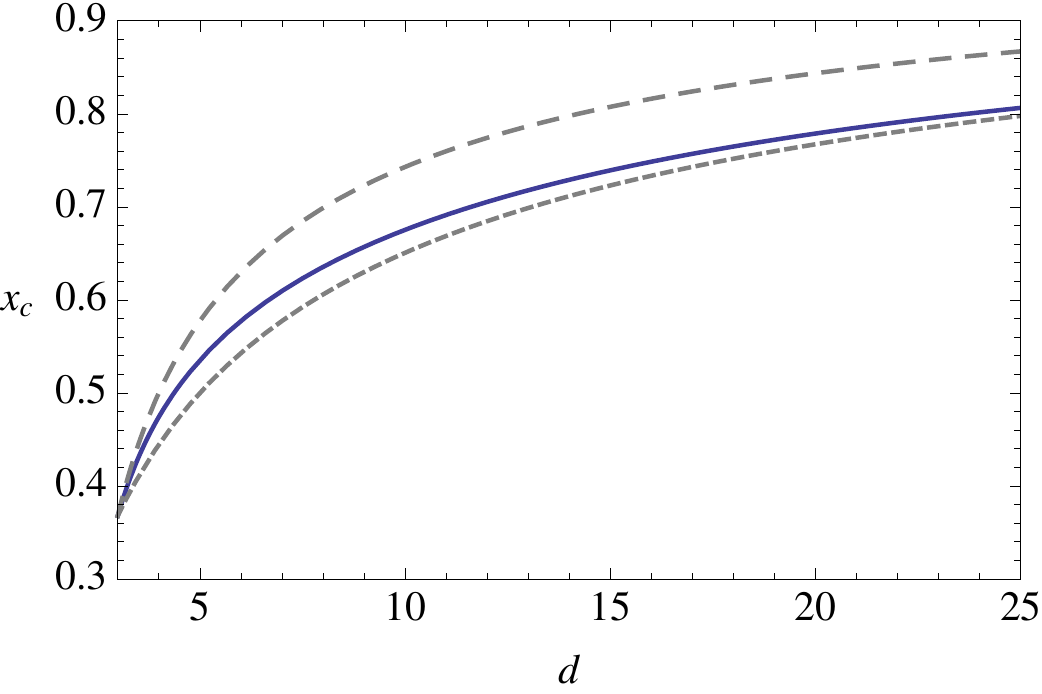}
\caption{\label{pxc} The ratio $x_c$ of the outer horizon to the classical Schwarzschild horizon for critical non-rotating black holes as a function of the number of dimensions based on  \eq{Gk} in the linear approximation (dashed line, $\delta=1$),  quadratic approximation (short dashed line, $\delta=2$), and with the full RG running \eq{delta} (solid line).}
\end{figure}

Degenerate (critical) black holes are achieved for $\Delta=0=\Delta'$ as this function is defined in \eqref{qualquant}. 
Using the expression \eqref{Gk} for the running of Netwton's coupling together with the above definitions, we obtain the critical values $x_c$ and $\Omega_c$ for non-rotating black holes as
\begin{eqnarray}
x_c&=&\left(\frac{\delta}{d-3+\delta}\right)^{\frac{\delta}{d-3}}\\
\Omega_c&=&\frac{d-3}{d-3+\delta}\left(\frac{\delta}{d-3+\delta}\right)^{\frac{\delta}{d-3}}\,.
\label{OmegaDelta}
\end{eqnarray}
Note that $\Omega_c=\frac{d-3}{d-3+\delta}\,x_c$. For $\delta=1$, they reduce to expressions first derived in \cite{kevin}. For $\delta$ as predicted by \eq{delta}, the result is displayed in Fig.~\ref{pxc}, also comparing the linear and quadratic approximations $\delta=1$ and $2$, respectively. We note that $x_c,\Omega_c\to 1$ with increasing dimensions. In the limiting case $d\to 3$, we have $x_c=\exp(-1)$ and $\Omega_c=0$. 
With these findings, the meaning of the parameter $\Omega$ becomes clear, and we write it as
\begin{equation}\label{omega}
\Omega=\left(\frac{M_c}{M}\right)^{\frac{d-2}{d-3}}\,\Omega_c
\end{equation}
with $\Omega_c$ given by \eq{OmegaDelta}. The significance of \eq{omega} is that black hole solutions exist for black holes masses down to the critical mass $M=M_c$, but not below. The mass scale  $M_c$ does not exist within the classical framework and is the central new ingredient here. It is expressed as
\begin{equation}
\frac{M_c}{M_P}=\left(g_*\,\Omega_c\right)^{-\frac{d-3}{d-2}}\xi^{d-3}.
\end{equation}
in terms of the Planck mass and the parameters of the RG. Below, we use the critical mass of the (non-rotating) black hole as a reference scale for the analysis of the rotating black holes.

In the sequel, it is often sufficient to use the approximation $\delta=1$, in which case the expressions for $\tilde \Delta$ and $\Omega$ become
\begin{eqnarray}
\tilde\Delta&=&A+x^2-\frac{x^3}{x^{d-2}+\Omega}
\label{correctedDelta}.\\
\Omega&=& (d-3)(d-2)^{-\frac{d-2}{d-3}}\left(\frac{M_{\textnormal{c}}}{M}\right)^{\frac{d-2}{d-3}}
\label{OmegaEx}
\end{eqnarray}
These equations and their solutions are the subject of the following sections.

\subsection{Four dimensions}

\begin{figure*} 
\centering
\includegraphics[width=.6\hsize]{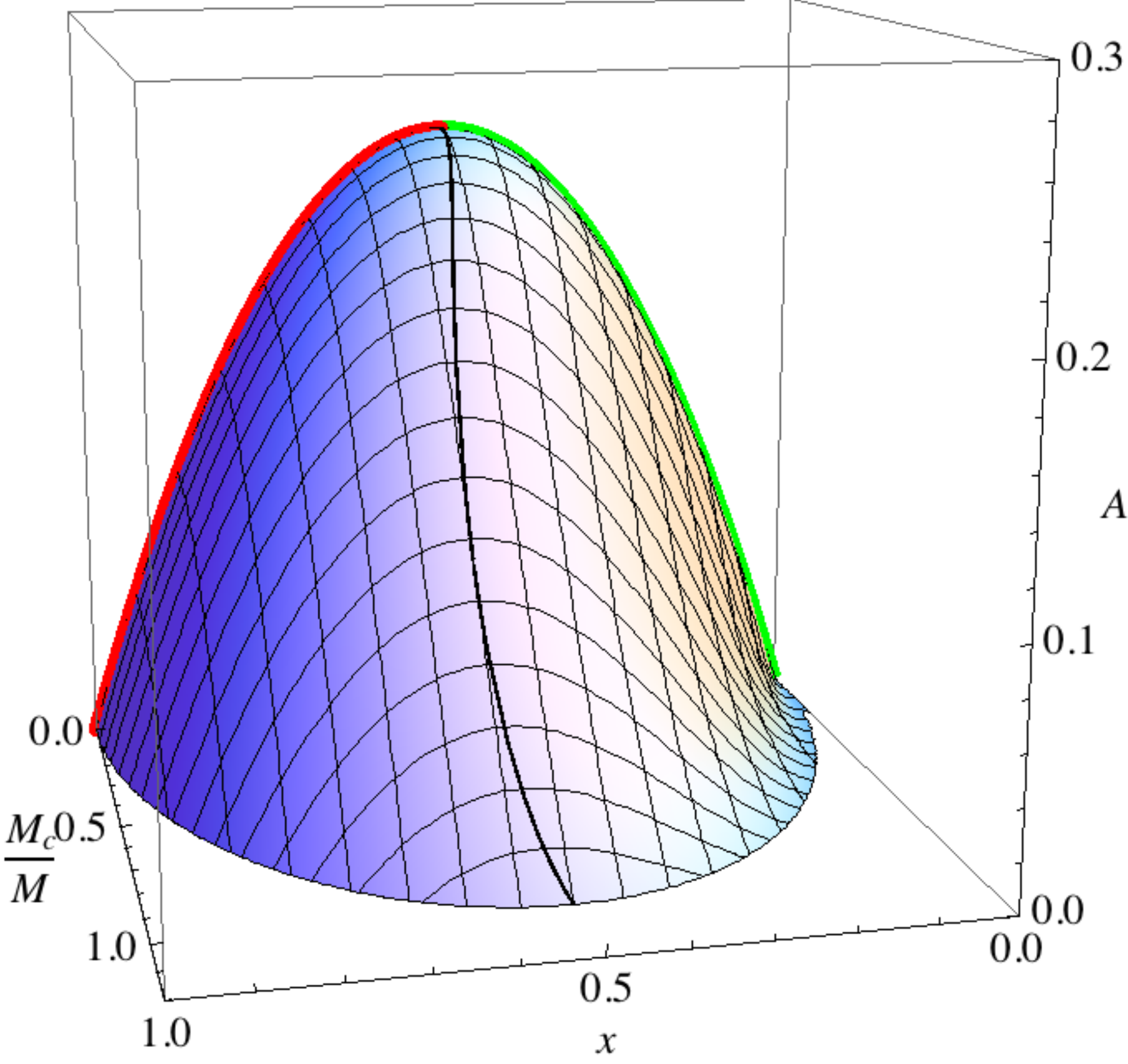}
\caption{\label{WXA4}{\small The phase space of black hole solutions in four dimensions. The points on the two-dimensional surface represent horizon radii $x$ \eq{x} as a function of angular momentum $A$ \eq{A} and the mass ratio $M_{\textnormal{c}}/M$. The thick black line identifies the radius of the critical horizon $x_{\textnormal{c}}(A,M;M_c)$. For fixed mass, the regions with $x>x_{\textnormal{c}}(A)$ ($x<x_{\textnormal{c}}(A)$) correspond to the event (Cauchy) horizons. The red (green) lines represent the classical solutions (Cauchy) horizon, respectively,  in the limit $M_c/M\to 0$.}}
\end{figure*}

\begin{figure*} 
\centering
\includegraphics[width=.6\hsize]{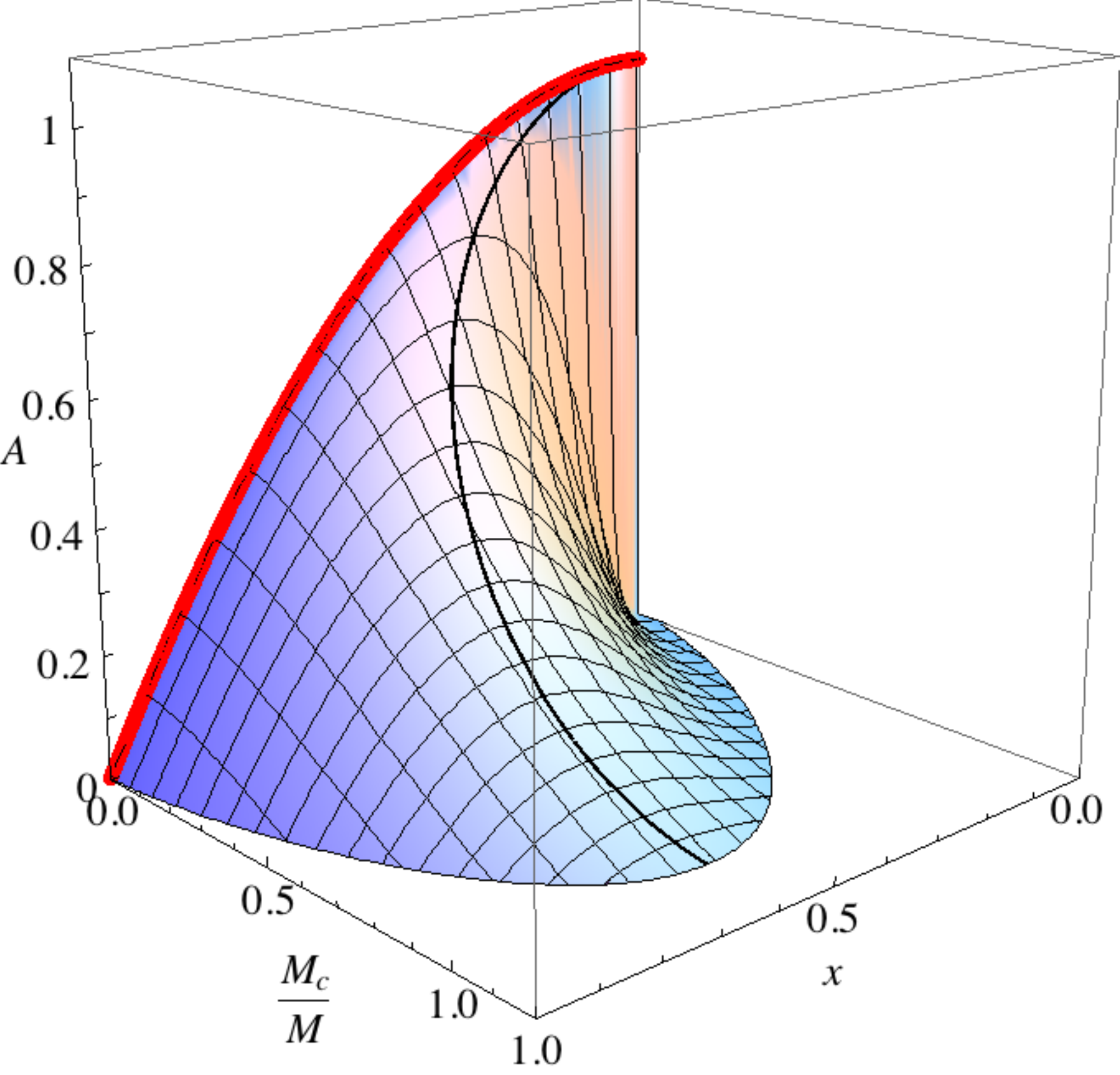}
\caption{\label{WXA5}{\small The phase space of black hole solutions in  five dimensions. The points on the two-dimensional surface represent the horizon radii $x$ as a function of angular momentum $A$ and mass $M_{\textnormal{c}}/M$. The thick black line gives the radius of the critical horizon $x_{\textnormal{c}}$. The regions with $x>x_{\textnormal{c}}(A)$ ($x<x_{\textnormal{c}}(A)$) correspond to event (Cauchy) horizons. The classical solution in the limit $M_c/M\to 0$ only admits an event horizon (red line).}}
\end{figure*}

The  RG-improved Kerr solution has been studied  in \cite{tuiran}. We recall this case for completeness, and in order to compare with the higher-dimensional results. 
Classical Kerr black holes in four dimensions posses two horizons (see Sec.~\ref{review}), corresponding to 
$A\leq \frac{1}{4}$. Concequently, for every angular momentum $J$ there is a minimum mass $M_{\textnormal{c}}(A)$, bellow which there are no classical Kerr solutions. This structure remains unchanged even under the inclusion of RG corrections \cite{tuiran}, except that the precise bounds depend, additionally, on the RG parameter $\Omega$. 
 Specifically,  by solving simultaneously $\tilde \Delta=0$ and $\tilde \Delta'\ge0$, we find a relation
between the permitted values of $A$ and $\Omega$,
\begin{eqnarray}
0&\le & \frac{3}{32}+\frac{1}{32}\sqrt{9-32(A+\Omega)}\nonumber\\
&&
-\frac{1}{3}(A+\Omega)
-\frac{1}{6}\sqrt{A^2+\Omega^2+10A\Omega}\,.\label{condition4}
\end{eqnarray}
Inspection of the above condition shows that as $\Omega$ increases the upper limit of $A$ decreases. This means, that as the mass of the black hole gets smaller and quantum effects become important, the maximally allowed value of the ratio $A$ decreases. This is in contrast to the classical picture where the condition $A\leq\frac{1}{4}$ was sufficient for the existence of black holes at all mass scales. The quantum corrected picture implies that for every black hole mass there is a different bound of the angular momentum parameter $A_{\textnormal{c}}(M)$ as an upper limit in order to have horizons. When we go down to the critical mass for non rotating black holes $M_{\textnormal{c}}$ we find that the angular momentum parameter should vanish. For masses less than this there is no allowed phase space for black holes. The horizons and the allowed phase space of black hole solutions can be seen in Fig.~\ref{WXA4}. For convenience, in Figs.~\ref{WXA4},~\ref{WXA5}, and \ref{WXA6} we have displayed the ratio $M_c/M$  instead of $\Omega$, using \eq{OmegaEx}.

We can also view the criticality condition in the opposite way. It is evident from \eqref{condition4} that for every value of the ratio $A\leq\frac{1}{4}$, there is a maximum allowed value of $\Omega$ for which black hole solutions exist. This value corresponds to the minimum mass $M_{\textnormal{c}}(A)$.
Moreover, we can deduce from \eqref{condition4} that as the angular momentum parameter $A$ grows the minimum required mass for the existence of horizons $M_{\textnormal{c}}(A)$ increases. For the classically critical black hole with $A=\frac{1}{4}$ we find that $\Omega=0$ and so that $M_{\textnormal{c}}(A=\frac{1}{4})\to\infty$ implying that only macroscopic black holes can reach this limit.

The vanishing of \eqref{condition4} defines the relation between $A$ and $\Omega$ when we are at criticality. Then, we can find the radius of the critical horizon by solving $\tilde \Delta(x)=0$ and $\tilde \Delta'(x)=0$ simultaneously. This gives
\begin{equation}
x_{\textnormal{c}}=\frac{3}{8}+\frac{1}{8}\sqrt{9-32(A+\Omega)},\label{xcr4}
\end{equation}
where it should be kept in mind that $A$ and $\Omega$ are implicitly related through the vanishing of \eqref{condition4} and the radius of the critical horizon is a function of only one parameter, $x_{\textnormal{c}}(A)$ or $x_{\textnormal{c}}(\Omega)$. This reflects the two directions of criticality in rotating black holes.
After some analysis we find that the value of $x_{\textnormal{c}}$ for every possible $A$ and $\Omega$ ranges from $x_{\textnormal{c}}=0.5$ to $x_{\textnormal{c}}\simeq0.55$.

From Fig. \ref{WXA4} we can observe how the horizons vary when we change angular momentum $A$ and $\Omega$ (or $M_c/M$). As any of these two parameters grows, the radius of the event horizon decreases while the Cauchy horizon increases. Both horizons meet, when we have reached the extreme configuration, at the critical horizon $x_{\textnormal{c}}$. This behaviour is verified if we look at the variation of the roots with respect to $A$ and $\Omega$
\begin{equation}
\partial_{\Omega}x_{\pm}=-\frac{\partial_{\Omega}\tilde\Delta|_{x_{\pm}}}{\tilde\Delta'(x_{\pm})}, \qquad \qquad \partial_{A}x_{\pm}=-\frac{\partial_{A}\tilde\Delta|_{x_{\pm}}}{\tilde\Delta'(x_{\pm})} . \label{variationofroots}
\end{equation}
Since, $\partial_{\Omega}\tilde\Delta|_{x_{\pm}}$ and $\partial_{A}\tilde\Delta|_{x_{\pm}}$ are always positive while $\tilde\Delta'(x_{\pm})$ is positive at $x_+$ and negative at $x_-$, it is implied that  $\partial_{\Omega}x_{\pm}$ and $\partial_{A}x_{\pm}$ are negative at the event horizon and positive at the Cauchy horizon. Note that this result is independent of dimensionality and it is true for every $d$.

\subsection{Five dimensions}

Classical Myers-Perry black holes with $d=5$ constitute a marginal case withonly one horizon, provided an auxiliary condition between angular momentum and mass is satisfied. In terms of dimensionless variables this condition reads $A\leq1$. 

The effect of quantum corrections is both to alter the horizon structure and to modify the condition for the existence of horizons. The horizon structure is modified as soon as we leave the classical limit. This is seen from the equation \eqref{correctedDelta} which gives the horizons. As soon as $\Omega$ takes any non-zero value, space-time develops a second (Cauchy) horizon provided that solutions to \eqref{correctedDelta} exist. When we reach the critical black hole configurations these two horizons meet at $x_{\textnormal{c}}$. This structural change can be observed in Fig. \ref{WXA4}.

The condition for the existence of horizons, changes from being a simple bound on the ratio of angular momentum over the mass to a more complicated condition which depends on the mass scale of the black hole. We find this condition by solving simultaneously the two equations $\tilde \Delta(x)\leq0$ and $\tilde \Delta'(x)=0$. We obtain the following relation between the allowed values of $\Omega$ and $A$
\begin{equation}
\Omega\leq\frac{5-\sqrt{24A+1}}{1+\sqrt{24A+1}}\left(\frac{1}{6}-A+\frac{1}{6}\sqrt{24A+1}\right)^{3/2}\,.
\label{condition5}
\end{equation}
The relation reflects the two directions of criticality. For every angular momentum parameter $A\leq1$ there exist a minimum mass $M_{\textnormal{c}}(A)$ for which we can have black holes. As $A$ increases then $M_{\textnormal{c}}(A)$ grows and the classically critical black hole with $A=1$ can be reached only by macroscopic black holes, since in that case $M_{\textnormal{c}}(A=1)\to\infty$. Similarly, we can read the condition \eqref{condition5} as it defines the maximum allowed ratio $A_{\textnormal{c}}(M)$ for every mass. The horizons and the allowed phase space of five dimensional black holes are plotted in Fig.~\ref{WXA4}.

Solving for $\tilde \Delta(x)=0$ and $\tilde \Delta'(x)=0$ simultaneously we find that for the radius of critical horizon is given by
\begin{equation}
x_{\textnormal{c}}=\left(\frac{1}{6}-A+\frac{1}{6}\sqrt{24A+1} \right)^{1/2}.\label{xcr5}
\end{equation}
Similarly, the radius of the critical horizon can be expressed in terms of the mass scale $\Omega$ if we solve \eqref{condition5} for $A$ and substitue back to \eqref{xcr5}. In either case the critical horizon ranges from $x_{\textnormal{c}}=\sqrt{6}/4$ to $x_{\textnormal{c}}=0$.

We conclude that the `phase space' of quantum-corrected black holes in five dimensions is similar to the one in four. This can also be confirmed by studying  the equations \eqref{variationofroots}. As either $A$ or $\Omega$ increases the radius of the event horizon gets smaller, the radius of the Cauchy horizon grows and they meet when we reach $M_{\textnormal{c}}(A)$ or $A_{\textnormal{c}}(M)$.

\subsection{Six and more dimensions}\label{SixOrHigher}

Rotating six and higher-dimensional black holes offer a new feature in the classical case, because they display always one horizon without any restrictions on the angular momentum. As a consequence, there exist black holes with arbitrary large angular momentum, the so-called ultra-spinning black holes. Their horizon structure is then similar to that of a conventional Schwarzschild black hole.

The impact of quantum effects turns out to be more substantial in this case, altering both of these features. In the quantum-improved picture, six or higher dimensional black holes display two horizons rather than one (an event horizon and a Cauchy horizon), and  only if an auxiliary condition between mass and angular momentum is satisfied. An arbitrary small value of $\Omega$ implies that there is a maximum of the gravitational potential (see section \ref{generalremarks}) and so that there is a maximum value for the angular momentum parameter $A$. Note that this change of behaviour sets in as soon as smallest quantum corrections are admitted (e.g.~even for large black hole masses), as is seen from 
\eqref{correctedDelta}.

More specifically, we solve again the two relations $\tilde \Delta(x)\leq0$ and $\tilde \Delta'(x)=0$ to obtain the condition between $A$ and $\Omega$ for the existence of the horizons. However, for arbitrary dimensionality we get the expression
\begin{equation}
3-d+2\frac{A}{x_{\textnormal{c}}^2+A}-(d-2)\frac{\Omega}{x_{\textnormal{c}}^{d-2}+\Omega}\leq0
\end{equation}
where we have to keep in mind that $x_{\textnormal{c}}$ and $A$ are implicitly related through the equation
\begin{equation}
1+\frac{2A}{x_{\textnormal{c}}^2+A}-(d-2)x_{\textnormal{c}}^{d-5}(x_{\textnormal{c}}^2+A)=0
\end{equation}
\begin{figure}
\centering

\includegraphics[width=.6\hsize]{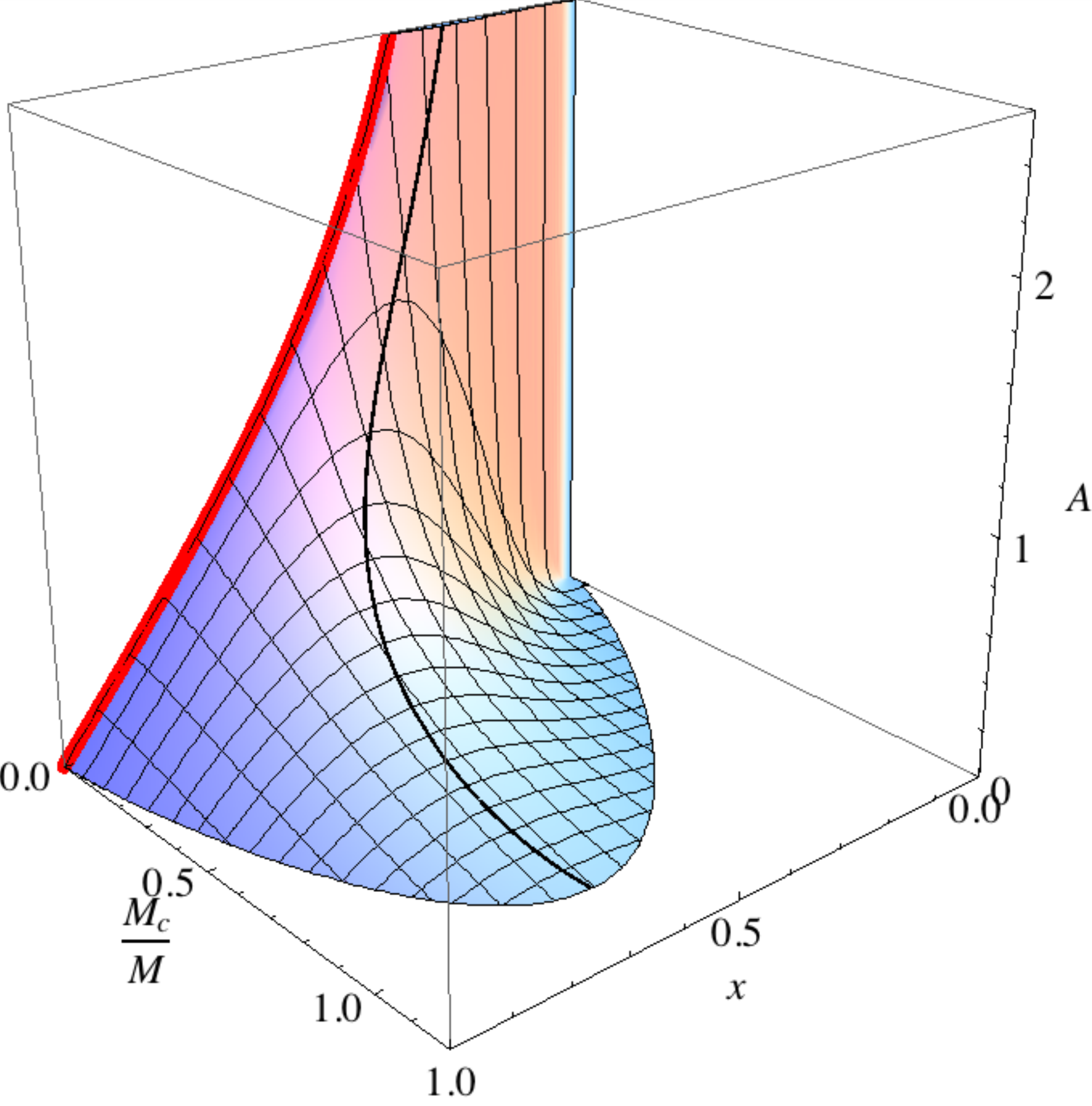}
\caption{\label{WXA6}{\small The allowed phase space of black hole solutions in six dimensions. The points of the two-dimensional surface represent the radius of the horizon $x$, as a function of $A$ and $M_{\textnormal{c}}$. The thick black line gives the radius of the critical horizon $x_{\textnormal{c}}$. The regions with $x>x_{\textnormal{c}}(A)$ ($x<x_{\textnormal{c}}(A)$) correspond to event (Cauchy) horizons. The red line at $M_c/M=0$ represents the classical event horizon. }}
\end{figure} 
A very interesting consequence of quantum effects is that they impose an upper bound in the angular momentum and in addition, that the smaller the black hole mass is, this maximum angular momentum gets smaller and smaller until it should vanish. This has as a result that ultra-spinning black holes do not exist in the presence of quantum effects. To see this more clearly, we make the approximation $\frac{A}{x}\gg1$ (which corresponds to the ultra-spinning regime) and we solve the two relations $\tilde \Delta(x)\leq0$ and $\tilde \Delta'(x)=0$ in order to find the condition for the existence of horizons in this regime. Then we get

\begin{equation}
\Omega\leq\frac{d-5}{3}\left[A\left(1+\frac{d-5}{3}\right)\right]^{-\frac{d-2}{d-5}}.\label{omegacr}
\end{equation}    

It is evident from the above inequality, that for $d\geq6$ and $\frac{A}{x}\gg1$ the maximally allowed value for $\Omega$ becomes extremely small as it scales as an inverse power of $A$. This means that ultra-spinning black holes exist only in the semi-classical regime, where the black hole mass $M\gg M_{\textnormal{c}}$ and therefore $\Omega\to0$. We can interpret this feature as an instability of ultra-spinning black holes under even small quantum fluctuations of the metric field. This behavior of  black holes in $d\geq6$ dimensions and their allowed phase space of solutions can be observed in Fig.~\ref{WXA6}.

The behaviour of the event and Cauchy horizons under variations of the parameters $A$ and $\Omega$ is the same as in the four and five dimensional case. This is verified by studying the equations \eqref{variationofroots}. Then, we observe that with increasing$A$ or $\Omega$ the event horizon shrinks, and the inner horizon grows until both meet for the critical values of the parameters $A_{\textnormal{c}}(M)$ or $M_{\textnormal{c}}(A)$ at the critical horizon.

\subsection{Ergosphere}

From the form of the corrected metric (\eqref{MP} with the substitution $G_N\to G(r)$) we observe that it still possesses the timelike Killing vector $k=\frac{\partial}{\partial t}$. The ergosphere is the region outside the event horizon, where $k$ becomes spacelike, 
\begin{equation}
k^{\mu}k_{\mu}=g_{tt}=\frac{a^2\sin^2\theta-\Delta(r)}{\Sigma(r)}>0\label{ergoregion}
\end{equation}
with $\Sigma(r)$ and $\Delta(r)$ defined by $\eqref{D}$.
This is a region where an observer cannot remain stationary. All observers in the ergosphere are forced to rotate in the direction of rotation of the black hole. It has been suggested that ergosphere can be used to extract energy from rotating black holes through the Penrose process \cite{Christodoulou:1970wf}, \cite{Christodoulou:1972kt}. 

The boundary of ergoregion is modified due to quantum corrections. 
First, note that we are interested in finding the roots of a new function $E(r)=a^2\sin^2\theta-\Delta(r)$, which is the original $\Delta(r)$ shifted by an angular-dependent term. Revisiting the analysis of Sec. \ref{generalremarks} we conclude that $E(r)$ has either two, one or no roots, depending on the values of its parameters. 
We denote (in dimensionless variables) the larger root of \eqref{ergoregion} as $x_{\small{E}+}$ and the smaller as $x_{\small{E}-}$. Then the ergoregion will be the region $x_+<x<x_{\small{E}+}$.
\begin{figure} [t]
\centering
\includegraphics[width=.7\hsize]{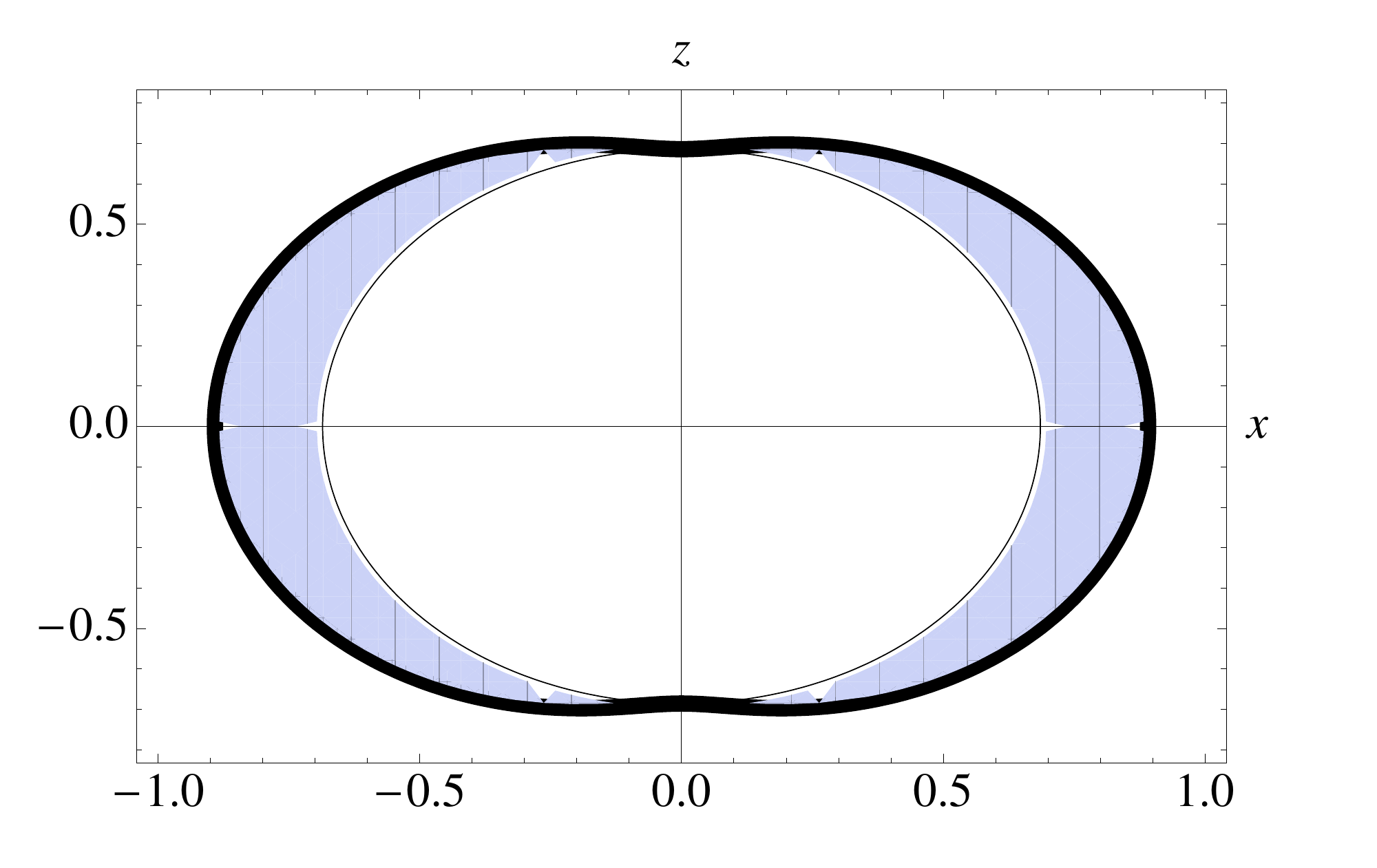}
\caption{\label{erg1}The ergosphere in six dimensions with $A=1/5$ and $\frac{M}{M_{\textnormal{c}}}=1.56$, plotted in the dimensionless $x-z$ plane with the angular coordinate $\theta$ starting from the axis of rotation $z$. The ergoregion is represented by the shaded region between the event horizon (inner solid line) and the boundary $x_{E+}$ (outer thick line).  
}
\end{figure} 
\begin{figure}
\vskip-.3cm
\centering
\includegraphics[width=.7\hsize]{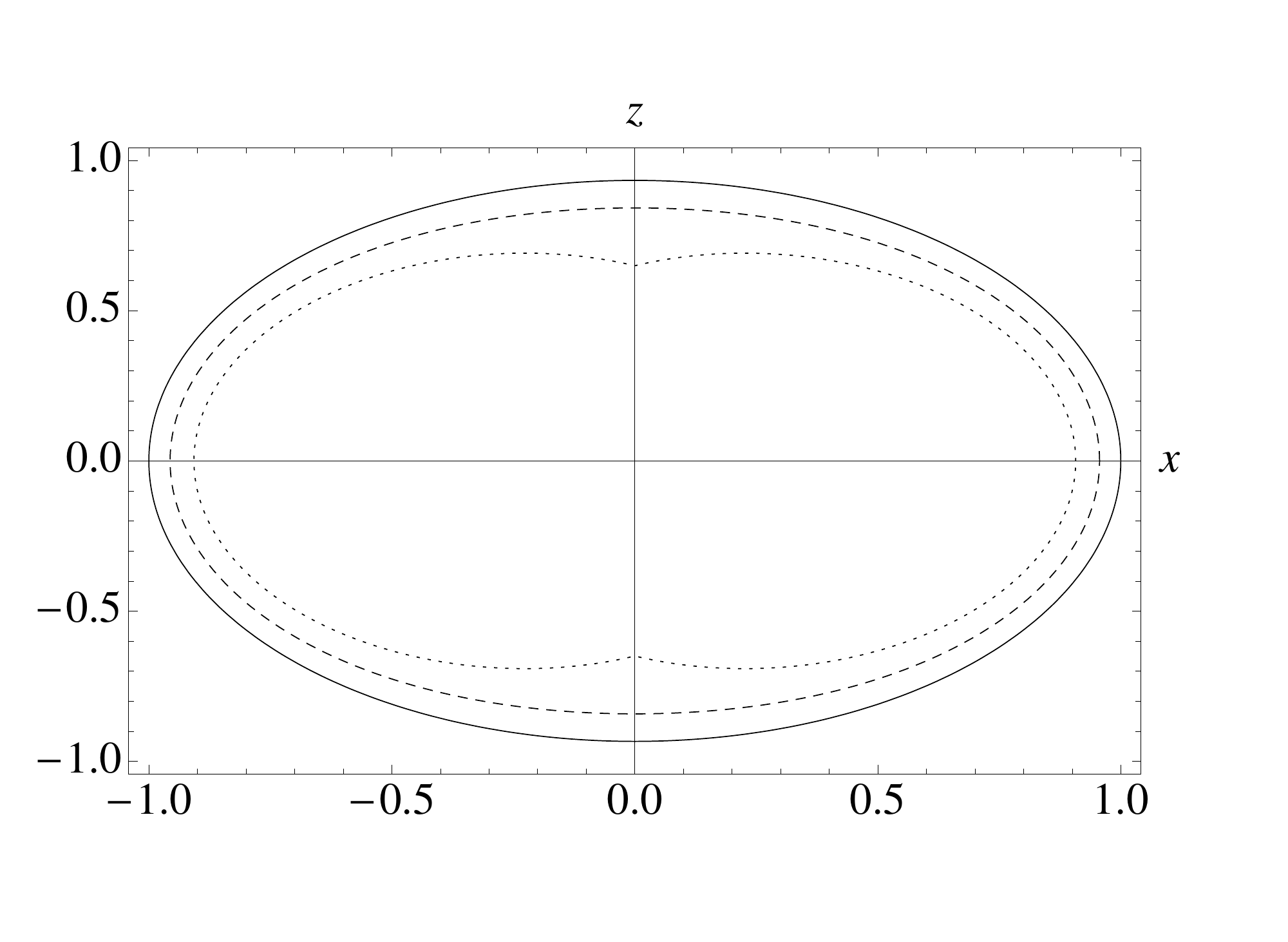}
\vskip-.5cm
  \caption{\label{erg2} The boundary of the ergoregion $x_{E+}$ for rotating black holes with $A=1/4$ in six dimensions for various black hole masses in the dimensionless $x-z$ plane. Classical, quantum, and critical black holes are represented  by a solid, dashed, and dotted line corresponding to the  parameters $\frac{M_{\rm c}}{M}\to 0$,  $\frac{M}{M_{\rm c}}=2.77$ and $\frac{M}{M_{\rm c}}=1.72$, respectively.}
\end{figure}
In order to examine the properties of $x_{\small{E}+}$, we write down the function $\tilde E(x)$ in terms of dimensionless variables
\begin{equation}
\tilde E(x)=-x^2-A\cos^2\theta+\frac{x^3}{x^{d-2}+\Omega}
\end{equation}
and we find its largest root. The form of this equation shows that the solution we are looking for will depend on the angular coordinate $\theta$. The analysis for $\tilde\Delta(x)$ is easily extended for $\tilde E(x)$ noticing that their only difference is the substitution of the angular momentum term by $A\cos^2\theta$. This means that $\tilde E(x)$ interpolates from the zero angular momentum $\tilde\Delta(x)|_{A=0}$ at the equatorial plane to the full $\tilde\Delta(x)$ at the poles. Moreover, we know from the analysis of $\tilde\Delta(x)$ that as the angular momentum parameter $A$ increases the largest root $x_+$ decrease. Thus, the outer boundary of ergoregion $x_{E+}$ coincides with the event horizon at the poles and as the angle $\theta$ grows, $x_{E+}$ grows until it reaches the horizon of the non-rotating limit at $\theta=\pi/2$. The outer boundary of the ergoregion $x_{E+}$ and the event horizon are plotted in the $x-z$ plane in Fig. \ref{erg1}.

Next, we answer the question wether  $x_{\small{E}+}$ is larger or smaller than the classical value. This is straightforward if we consult the analysis of the previous sections. Since $\tilde E(x)$ is just $\tilde\Delta(x)$ with a different angular momentum parameter, its behaviour with varying $\Omega$ is the same. That is, if we go to smaller masses and quantum effects become important (larger $\Omega$) then  $x_{\small{E}+}$ gets smaller. This can also be observed in Fig. \ref{erg2} where we have plotted  $x_{\small{E}+}$ for different values of $\Omega$.

Finally, we have to comment if any modifications to the structure of ergosphere are coming from different dimensions. It should be clear by now that this is not the case. The function $\tilde E(x)$ follows the behaviour of $\tilde\Delta(x)$ and it has always two, one or no horizons for every dimensionality. Since we are interested only for the larger root $x_{E+}$ of $\tilde E(x)$, it makes no difference if classically $\tilde E(x)$ had only one root (as is the case for $d\geq5$) or more (as in $d=4$). Moreover, as the angular momentum term in $\tilde E(x)$ is always less or equal to that of $\tilde\Delta(x)$, the condition for the existence of the event horizon is enough to guarantee the existence of the ergosphere. It also follows that if the function $\tilde E(x)$ will have a second root (as in the quantum corrected case) this will always be smaller than $x_-$ and thus will be irrelevant for the ergosphere, which is the region $x_+<x<x_{\small{E}+}$.

\section{Thermodynamics}\label{Thermodynamics}

The second part of our analysis will deal with the thermodynamical properties of the quantum corrected black holes. We present some basic properties of quantum-corrected Myers-Perry space-times, such as the angular velocity, the area and the surface gravity of the horizon. Then, we examine the temperature and the specific heat.

\subsection{Killing vectors}

For studying the thermodynamical properties of black holes we make use of the Killing vectors of the space-time. We observe from the corrected form of \eqref{MP} that we still have both Killing vectors present in the classical case, namely $k=\frac{\partial}{\partial t}$ and $m=\frac{\partial}{\partial\phi}$ associated with time translations and axisymmetry, respectively. . 
We note that the results of this section are applicable for a generic form of the function $G(r)$ which parametrizes the running of Newton's constant.

We begin by finding the null generator of the event horizon. 
In order to avoid a coordinate singularity at the horizon, we proceed with the coordinate transformation
\begin{equation}
du=dt+\frac{r^2+a^2}{\Delta}dr,  \qquad  \qquad d\chi=d\phi +\frac{a}{\Delta}dr.\label{kerrcoordinates}
\end{equation} 
In these coordinates the Killing vectors are $k=\frac{\partial}{\partial u}$ and $m=\frac{\partial}{\partial\chi}$. The vector field normal to a hypersurface $S=const.$ is given by $l=f(x)\left(g^{\mu\nu}\partial_{\nu}S\right)\frac{\partial}{\partial x^{\mu}}$, where $f(x)$ is an arbitrary function. Using $S=r-r_+$ we find the normal vector at the event horizon
\begin{equation}
l_+=\frac{a^2+r_+^2}{r_+^2+a^2\cos^2\theta}\cdot f(r_+)\cdot\xi
\end{equation}
where the vector $\xi$ is given by
\begin{equation}
\xi=\frac{\partial}{\partial u}+\frac{a}{a^2+r_+^2}\cdot\frac{\partial}{\partial\chi}\label{xi}.
\end{equation}
It is easy to verify that the normal vector $l_+$ is null ($l_+^2=0$) and that the vector $\xi$ is a Killing vector of the metric transformed by \eqref{kerrcoordinates}. As a result, we have that the event horizon is a Killing horizon of the Killing vector field $\xi$. 

\subsection{Angular velocity}
The angular velocity of the horizon $\Omega_H$ is found by comparing orbits of the Killing vector $k$ (which correspond to static particles), with orbits of the Killing vector $\xi$ (which generates the event horizon). We find
\begin{equation}
\Omega_H=\left.\frac{d\phi}{dt}\right|_{r=r_+}=\frac{a}{r_+^2+a^2}\,.
\end{equation}
The functional form of this quantity is identical to its classical counterpart \cite{MyersPerry}, except for the value of $r_+$ which is different from its classical value. 

\subsection{Horizon area}
An expression for the area of the horizon is given by the integral $\int\sqrt{g^{(d-2)}}d\theta d\phi d\Omega_{d-4}$, performed at the event horizon, with $g^{(d-2)}$ the metric which corresponds to the geometry of the horizon. We find
\begin{equation}
\mathcal{A}_H=r_+^{d-4}(r_+^2+a^2)\Omega_{d-2}.
\end{equation}
This expression is functionally  the same as the classical one upon replacing the classical horizon radius by $r_+$.

\subsection{Surface gravity}
For a Killing horizon of a Killing vector $\xi$, the surface gravity $\kappa$ is defined as
\begin{equation}
\xi^{\nu} \nabla_{\nu}\,\xi^{\mu}=\kappa\,\xi^{\mu}\label{sg}.
\end{equation}
After substituting $\xi$ as in \eqref{xi} and performing the algebra we find the surface gravity as
\begin{equation}
\kappa=\frac{\left.\partial_r\Delta\right|_{r=r_+}}{2(r_+^2+a^2)}=\frac{1}{2r_+}\left(\frac{2r^2_+}{r_+^2+a^2}+d-5+\eta(r_+)\right),\label{surfgrav}
\end{equation}
where $\eta(r_+)=-r_+\frac{G'(r_+)}{G(r_+)}$ is the anomalous dimension. The expression reduces to the classical one if, firstly, $r_+$ is replaced by the classical radius and, secondly, the term proportional to the anomalous dimension is dropped. 
We recall that \eqref{EtaCritical}
\begin{equation}
\eta(r_{\textnormal{c}})=5-d-2\frac{r_{\textnormal{c}}^2}{r_{\textnormal{c}}^2+a^2}\label{etacritical}
\end{equation}
 in the case of a critical black hole where $r_+=r_c$, independently of the specific RG running. We thus conclude that critical black holes have zero surface gravity. For future use, we note that $\eta(r_+)$, for a running Newton's constant given by \eqref{Gofr} is a monotonically increasing function of $g_*$, starting from \eqref{etacritical}, when we are at $M_{\textnormal{c}}(\alpha)$ and taking its maximum value  $\eta(r_+)=0$ in the classical limit.

\subsection{Temperature}

Using the techniques of quantum field theory in curved space-time, Hawking showed that black holes radiate like thermal objects with temperature $T=\kappa/(2\pi)$, where $\kappa$ is the surface gravity of the event horizon \cite{Hawking}. Another method for identifying the temperature with $\kappa/(2\pi)$ comes from Euclidean quantum gravity techniques \cite{Gibbons:1976ue} where an identification in imaginary time with period $\beta=2\pi/\kappa$ is required in order to produce a smooth Euclidean manifold. In either case it is confirmed that the black hole temperature is a property of the space-time itself, independent of which theory of gravity determines the geometry \cite{Birrell:1982ix}, \cite{Jacobson:1993vj}.

Thus, it is straightforward to confirm that also in our case the formula for the temperature of the improved black holes is obtained by dividing the surface gravity \eqref{surfgrav} by $2\pi$,   
\begin{equation}
T=\frac{\kappa}{2\pi}=\frac{1}{4\pi r_+}\left(\frac{2r^2_+}{r_+^2+a^2}+d-5+\eta(r_+)\right).\label{temperature}
\end{equation}
The temperature depends on the horizon radius, angular momentum, dimensionality, and the graviton anomalous dimension.
Recalling that $r_+=r_+(M, J, g_*)$, we note that $T$  also depends, albeit implicitly, on mass and $g_*$.

From the expression \eqref{surfgrav} for the surface gravity and the limit values of $\eta(r_+)$, we see that for large enough masses the temperature will be positive. If there exists an extremal black hole (i.e. $\Delta'(r_+)=0$), then when mass gets its critical value $M_{\textnormal{c}}(a)$, the temperature vanishes. If the space-time does not exhibit an extremal solution, then the temperature remains positive, and it diverges as $r_+$ tends to zero. 

\begin{figure}
 \includegraphics[width=.65\hsize]{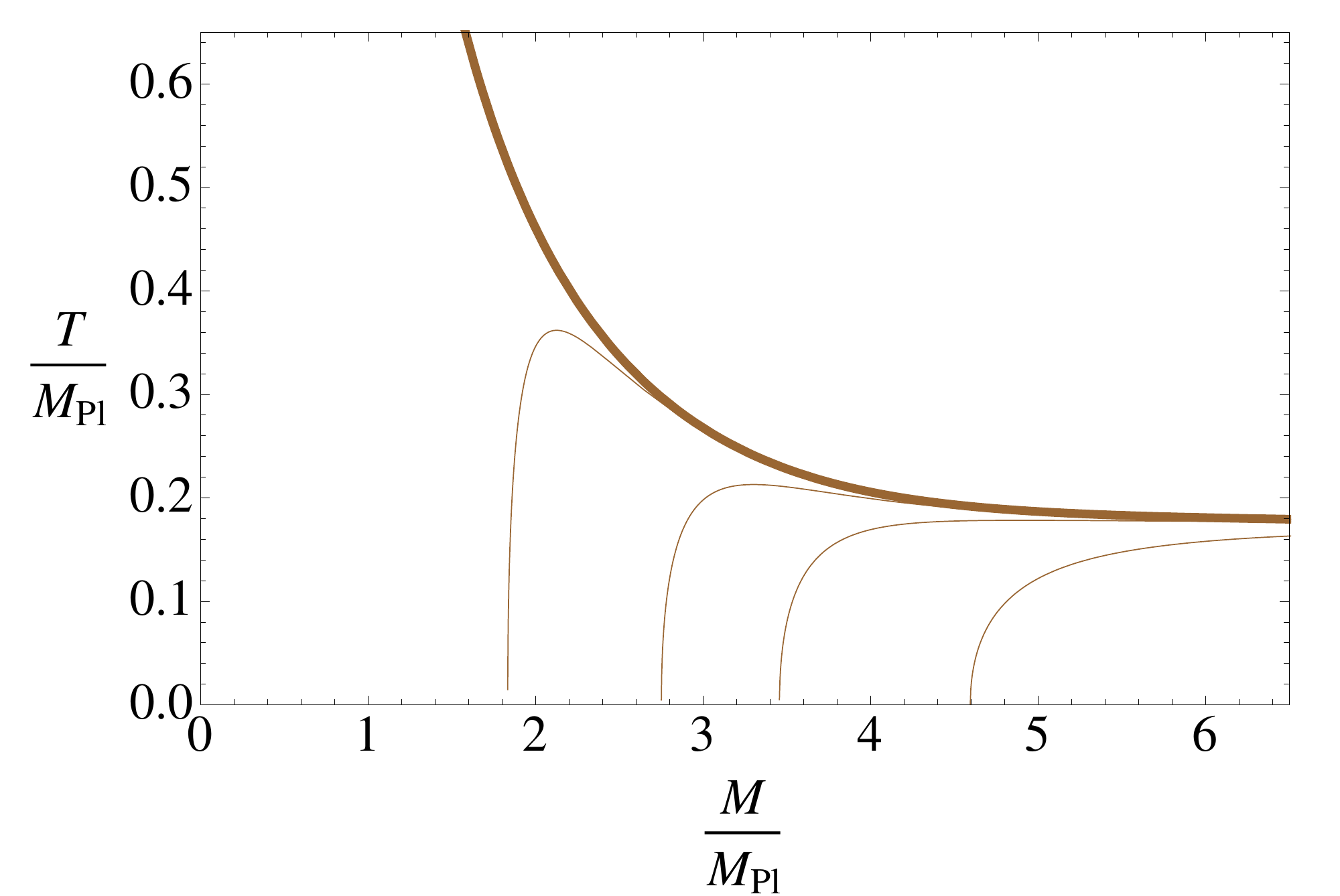}
 \caption{\label{TempW} The temperature for various values of $M_{\textnormal{c}}(a)$ in $d=7$ and fixed angular momentum $J=10$. The thick line corresponds to the classical temperature, while the other lines correspond to (from left to right) $M_{\textnormal{c}}(a)/M_{\textnormal{Pl}}=1.83$, $M_{\textnormal{c}}(a)/M_{\textnormal{Pl}}=2.75$, $M_{\textnormal{c}}(a)/M_{\textnormal{Pl}}=3.45$ and $M_{\textnormal{c}}(a)/M_{\textnormal{Pl}}=4.6$. We can verify that for  smaller $g_*$ (larger $M_{\textnormal{c}}(a)$) the temperature gets smaller. }
\end{figure}

We saw in the previous sections that four and five dimensional black holes exist only up to a critical mass $M_{\textnormal{c}}(a)$, both in the classical and the quantum regimes. Their temperature always reaches a maximum and vanishes when their mass reaches $M_{\textnormal{c}}(a)$. 
However, in six or higher dimensions there are no extreme configurations in the classical limit and the temperature diverges as $r_+\to0$. The picture is modified once quantum corrections are taken into account, since then black holes only exist up to a critical value $M_{\textnormal{c}}(a)$. Consequently, the temperature reaches a maximum. This change of behaviour can be seen in Fig. \ref{TempW} where the temperature in seven dimensions is plotted both for the classical and the improved black holes.

It is instructive to see how the temperature varies with the mass and $g_*$. We begin with the variation with respect to $g_*$. As explained before, this corresponds to changing the critical mass $M_{\textnormal{c}}(a)$ at fixed angular momentum $J$, given by the relation
\begin{equation}\label{dTdg}
\frac{d\,T}{d\,g_*}=
\frac{1}{4\pi}\left[\left(\frac{\Delta''(r_+)}{r_+^2+a^2}-\frac{2r_+\Delta'(r_+)}{(r_+^2+a^2)^2}\right)\frac{\partial r_+}{\partial g_*}-\frac{1}{r_+}\frac{\partial \eta(r_+,g_*)}{\partial g_*}\right]\,,
\end{equation}
Note that $\eta(r_+,g_*)$ depends both explicitly on $g_*$ due to the RG flow, and implicitly, via $r_+$. Using \eqref{Gofr}, we find that \eq{dTdg} is always negative. Moreover, the same result holds true for more general matchings of the form $k\sim r^{-\gamma}$ as long as the parameter $\gamma$ obeys $\gamma>(\sqrt{2}-1)/(2d-4)$. Thus, for fixed mass and angular momentum a larger value for $M_{\textnormal{c}}(a)$ implies a smaller black hole temperature. Consequently,  the temperature is always smaller than the classical one, i.e. $T_{\textnormal{cl}}(M,J)>T_{g_*}(M,J)$. We can observe the dependence of temperature to the critical mass $M_{\textnormal{c}}(a)$ by looking at the Fig. \ref{TempW}.

\begin{figure}[t]
\centering
\includegraphics[width=.7\hsize]{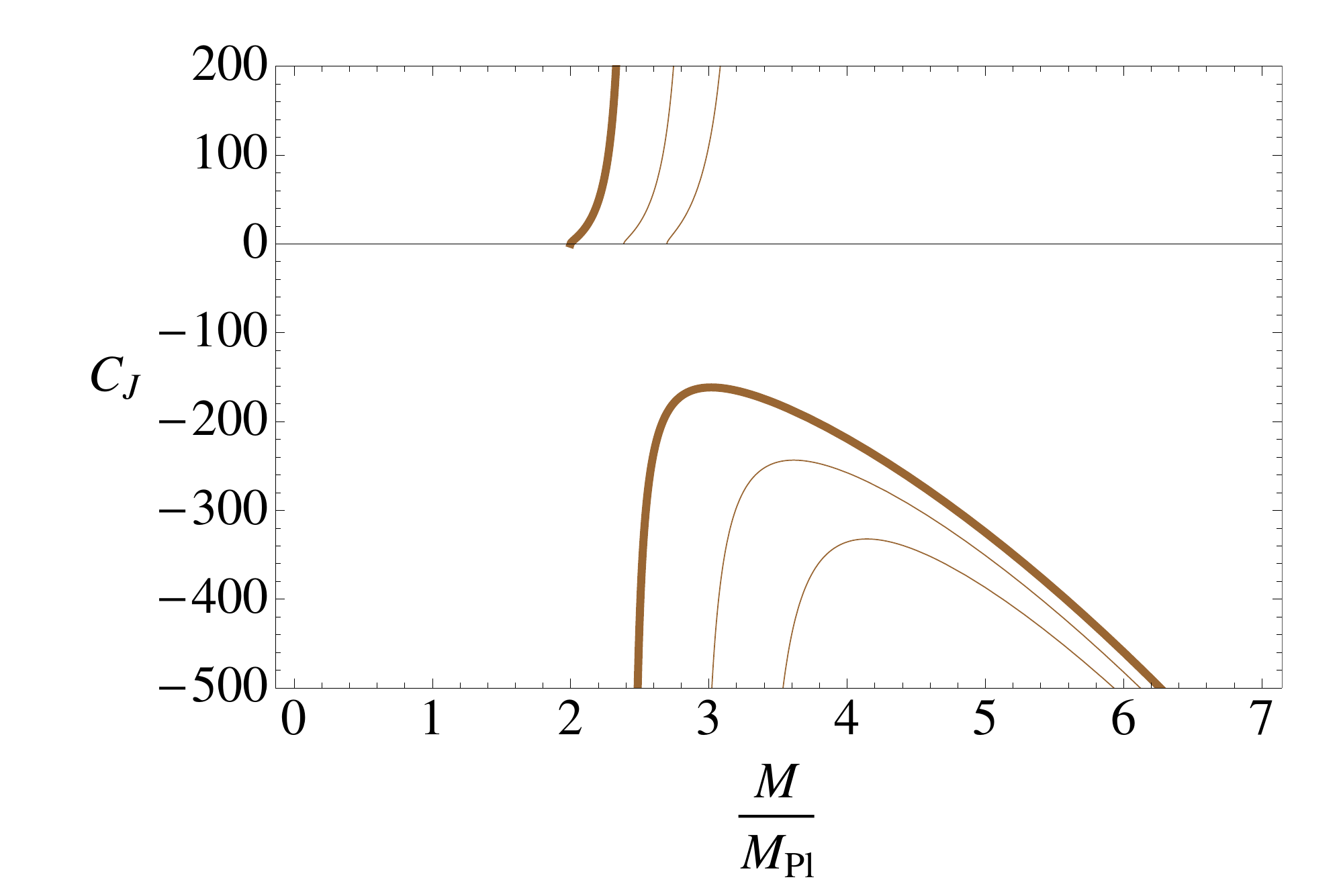}
  \caption{\label{CH4} {\small The specific heat for various values of $M_{\textnormal{c}}(a)$ in $d=4$ and fixed angular momentum $J=10$.  The thick line corresponds to the classical specific heat, while the thin lines correspond to (from left to right) $M_{\textnormal{c}}(a)/M_{\textnormal{Pl}}=2.4$ and $M_{\textnormal{c}}(a)/M_{\textnormal{Pl}}=2.7$. We observe that due to the fact that both in the classical and quantum cases there exist an extreme black hole there is no change in the qualitative behaviour of $C_J$.}}
\end{figure}

\begin{figure}[t]
\centering
\includegraphics[width=.7\hsize]{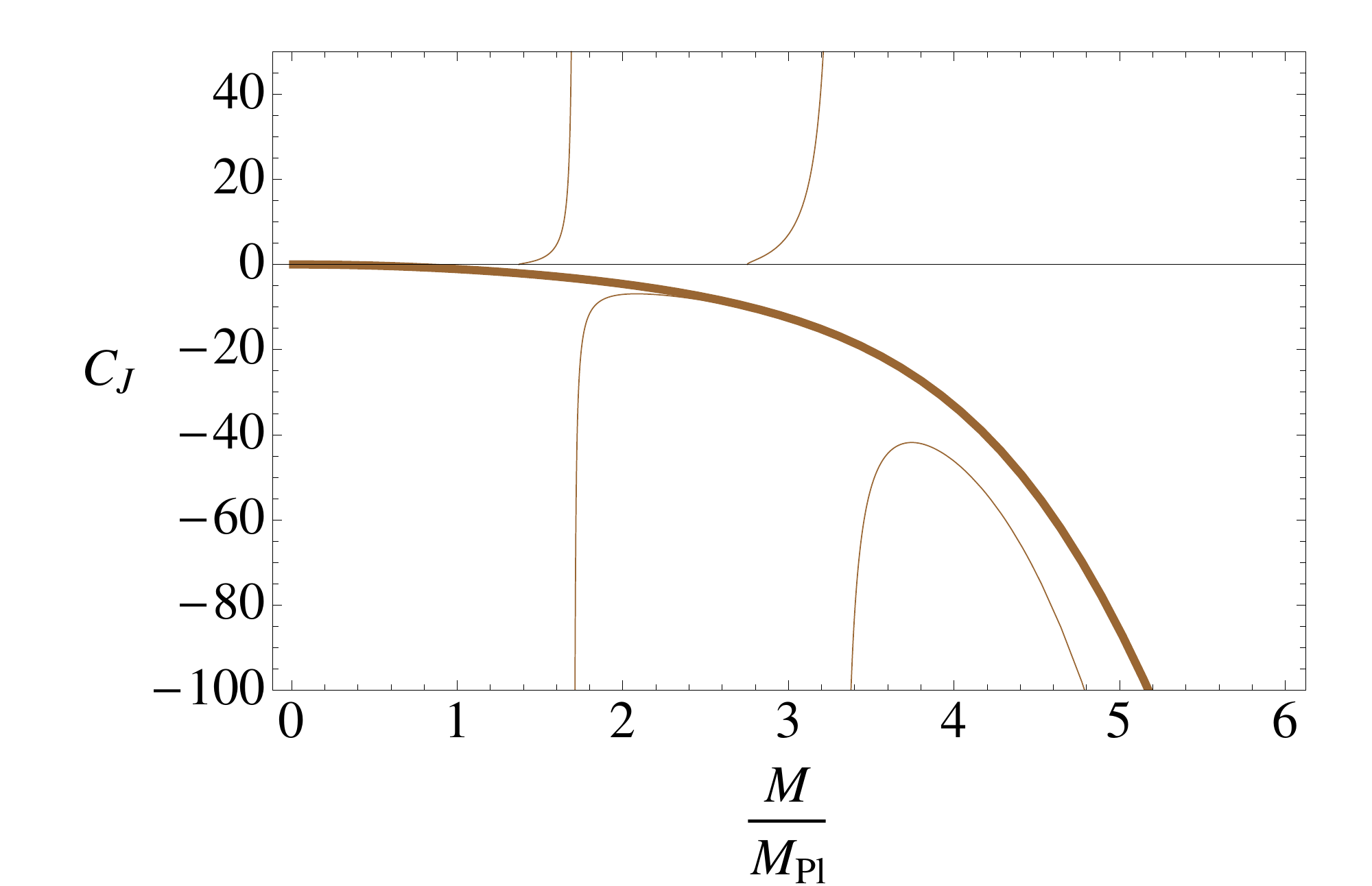}
  \caption{\label{CH7} {\small The specific heat for various values of $M_{\textnormal{c}}(a)$ in $d=7$ and fixed angular momentum $J=10$.  The thick line corresponds to the classical specific heat which is always negative, while the thin lines correspond to (from left to right) $M_{\textnormal{c}}(a)/M_{\textnormal{Pl}}=1.83$, $M_{\textnormal{c}}(a)/M_{\textnormal{Pl}}=2.75$ and have a pole where $C_J$ becomes positive for small masses.}}
\end{figure}

\subsection{Specific heat}

The specific heat at constant angular momentum is defined as
\begin{equation}
C_J=\left.\frac{\partial M}{\partial T}\right|_J.
\end{equation}
We are particularly interested in the sign of this quantity, since a positive specific heat implies a thermodynamically stable system. After some algebra we find
\begin{equation}
C_J=-(d-2)\,\pi\,\Omega_{d-2}\,G(r_+)\,\frac{ r_+^{d-3}(a^2 + r_+^2)^3\,T}{D(r_+)},
\end{equation} 
where $T$ is the temperature, and $D(r_+)$ is given by

\begin{widetext}
\begin{equation}
\begin{split}
D(r_+)=& \left[3(d-5)a^4- 6 a^2 r_+^2 + (d-3) r_+^4\right]G(r_+)^2 \\
 &-r_+^2 (a^2 + r_+^2) (3 a^2 + r_+^2) 
G'(r_+)^2  \\
 &+ r_+^2 G(r_+)\left[4 a^2 r_+ 
G'(r_+) + (a^2 + r_+^2) (3 a^2 + r_+^2) G''(r_+) \right]\,.\label{SHdenom}
\end{split}
\end{equation}
\end{widetext}
In the classical limit we have $G'(r_+)=0=G''(r_+)$ and the last two terms of \eqref{SHdenom} vanish. The fact that the temperature reaches a maximum and then vanishes at $M_{\textnormal{c}}$ in four and five dimensions is reflected by a pole and a change of sign in the specific heat (see Fig. \ref{CH4}). For six or higher dimensions the temperature is a monotonically decreasing function of mass, and the specific heat remains always negative (see Fig. \ref{CH7}).

When quantum corrections are considered in four and five dimensions the qualitative behaviour of the specific heat remains unchanged. However, the point where the specific heat becomes positive is shifted towards larger masses. For six or higher dimensions the effect of the quantum correction terms in \eqref{SHdenom} is that they induce always one pole at $C_J$ for some value of the mass. For masses sufficiently small, the specific heat becomes positive and vanishes at $M_{\textnormal{c}}(a)$. It is seen from \eqref{SHdenom} that the larger the value of $M_{\textnormal{c}}$ is, the pole is shifted towards larger masses.

\section{Mass and energy}\label{Mass and energy}

\subsection{Energy momentum tensor}

Myers-Perry black holes represent vacuum solutions of rotating space-times in higher dimensions. However, when we consider quantum corrections, the resulting space-times are no longer vacuum solutions of Einstein's equations, and we can think of them as arising from an effective energy momentum tensor $T_{\mu\nu}^{\textnormal{(eff)}}$. We find this effective enegy-momentum tensor by substituting the improved metric into Einstein's equation $G_{\mu\nu}=8\pi G_N T_{\mu\nu}^{\textnormal{(eff)}}$, leading to
\begin{equation}
T^{\mu\textnormal{(eff)}}_{\nu}=
\begin{pmatrix}
T^{t}_t & 0 & 0 & T^{t}_{\phi} & 0 & \cdots & 0 \\
0 & T^{r}_{r} & 0 & 0 & \vdots & & \vdots \\
0 & 0 & T^{\theta}_{\theta} & 0 & \vdots & & \vdots \\
T^{\phi}_{t} & 0 & 0 & T^{\phi}_{\phi} & 0 & & \vdots\\
0 & \cdots & \cdots & 0 & \ddots & 0 & \vdots\\
\vdots & & & & 0 & T^{i}_{i} & 0\\
0 & \cdots & \cdots & \cdots & \cdots & 0 & \ddots \label{EMtensor}
\end{pmatrix} 
\end{equation}
 where $4\leq i \leq d-1$ label the extra dimensions. We write a general entry of $T^{\mu\textnormal{(eff)}}_{\nu}$ as 
\begin{equation}
T^{\mu}_{\nu}=U^{\mu}_{\nu}G'(r)+V^{\mu}_{\nu}G''(r),\label{UandV}
\end{equation}
where in the classical limit $G'(r)=0=G''(r)$ and $T^{\mu}_{\nu}=0$ as expected. Then, we calculate the components and we find the coefficients $U^{\mu}_{\nu}$ and $V^{\mu}_{\nu}$. Their values are given in Appendix \ref{Appendix2}.
In order to examine the properties of the energy-momentum tensor we diagonalize \eqref{EMtensor}, finding
\begin{equation}
T=\textnormal{diag}(p_i), \qquad i=\{0,\cdots,d-1\}
\end{equation}
where $p_0\equiv-\rho$ is minus the energy density, $p_1=p_r$ and $p_2=p_3=p_{\perp}$, and $p_i$ for $i\ge 4$ arise from the higher dimensions. The energy density $\rho$ and $p_3$ arise after diagonalisation  of \eqref{EMtensor} while $p_r$ and $p_2$ originate directly from the $T^r_r$ and $T^{\theta}_{\theta}$. We refer to the Appendix \ref{Appendix2} for explicit expressions of these quantities. Note that $\rho=-p_r$ and that $p_2=p_{\perp}=p_3$.

\subsection{Energy conditions}
In classical general relativity many properties of the space-time depend on the energy conditions which are expressed as inequalities between the components of the energy momentum tensor. For this reason we compute the following relations relevant to these conditions:
\begin{eqnarray}
\rho\geq0:\mspace{2mu}\qquad\qquad\qquad\qquad\qquad\qquad\left[(d-2)r^2+(d-4)a^2\cos^2\theta\right]G'(r)\geq0\,\,\,\,&\label{energycondition0}\\
\rho+p_{\perp}\geq0:\quad\left[(d-2) r^2+(d-6)a^2\cos^2\theta\right] G'(r)-r \left(r^2+a^2 \cos^2\theta\right) G''(r)\geq0\,\,\,\,&\label{energycondition1}\\
\rho+p_i\geq0:\quad\left[(d-2) r^2+(d-4) a^2\cos^2\theta\right] G'(r)-r \left(r^2+a^2 \cos^2\theta\right) G''(r)\geq0 \,\,\,\,& \label{energycondition2}\\
\rho-p_{\perp}\geq0:\mspace{1mu}\,\,\quad\qquad\qquad\qquad\qquad\qquad\qquad\qquad\qquad(d-2)G'(r)+r\,G''(r)\geq0\,\,\,\,&\label{energycondition3}\\
\rho-p_i\geq0:\quad\left[(d-2) r^2+(d-4)a^2 \cos^2\theta\right] G'(r)+r \left(r^2+a^2 \cos^2\theta\right) G''(r)\geq0\,\,\,\,&  \label{energycondition4}
\end{eqnarray}
with the index $i$ being equal or greater than $4$. The sign of the above inequalities and consequently the validity of the energy conditions depend on the running of Newton's coupling through its first and second derivative. Evidently, in the absence of RG effects the effective energy-momentum tensor vanishe identically.

We now turn to the energy conditions. For a diagonalised energy-momentum tensor the \emph{weak energy condition} reads
\begin{equation}
\rho\geq0 \qquad  \textnormal{and}  \qquad \rho + p_{i} \geq0,\qquad 1\leq i\leq d-1.
\end{equation}
From the relation \eqref{energycondition0} we conclude that the sign of $\rho$ depends only on the sign of $G'(r)$ and in our case we have that $G'(r)>0$ so that the first requirement of the weak energy condition is always satisfied. Moreover, as stated above we have that $\rho+p_r=0$ and so for the validity of the weak energy condition we have to examine the remaining two relations $\rho+p_{\perp}\geq0$ and $\rho+p_i\geq0$ with $i\geq4$. The first of them is given by \eqref{energycondition1} and for a function $G(r)$ given by \eqref{Gofr} this expression has one root (denoted by $r_{w1}$). The condition is violated for $0< r < r_{w1}$. It is interesting to note that in the limit of zero angular momentum this condition is not violated. Moreover, the same holds true in the special case where $\theta=\pi/2$. The second relation we need to examine takes the form \eqref{energycondition2}. This relation has exactly the same behaviour as $\rho+p_{\perp}$ and is violated for $0<r<r_{w2}$ only when $a\neq0$ and $\theta\neq\pi/2$. Here, $r_{w2}$ denotes the root of the expression on the LHS of \eqref{energycondition2}.

Next we examine the validity of the \emph{dominant energy condition} which reads
\begin{equation}
 \rho\geq0 \qquad \textnormal{and} \qquad -\rho\leq  p_{i} \leq \rho ,\qquad 1\leq i\leq d-1.
\end{equation}
For this condition to hold, the weak energy condition should be fulfilled. We saw that there exist special cases for this to be true. So, now we have to examine the relations $\rho-p_{i}\geq0$. Again, the condition $\rho-p_r\geq0$ is always satisfied. The requirement that $\rho-p_{\perp}\geq0$ takes the form \eqref{energycondition3} and for $G(r)$  given by \eqref{Gofr} is violated for $r>r_{d1}$, where $r_{d1}$ is the only root of this expression. The remaining relations of the dominant energy condition are $\rho-p_i\geq0$ with $i\geq4$, and they take the form of \eqref{energycondition4}. However, for the only case where the weak energy condition is satisfied ($a=0$ or $\theta=\pi/2$), the above condition reduces to \eqref{energycondition3} and does not give any new information.

\subsection{Curvature singularities}

In classical general relativity, singularity theorems \cite{Hawking:1973uf} state that whenever an event horizon is formed, a curvature singularity is hidden behind this horizon. The derivation relies on the positivity conditions of any energy momentum present in the space-time. The fact that the RG-improved black holes violate some of these conditions opens the possibility that the space-time may no longerexhibit a singularity. 

Here we briefly comment on the fate of the singularities in our improved space-time. We compute two curvature invariants, the Ricci scalar $R$ and the Kretschmann invariant $K=R_{\mu\nu\rho\sigma}R^{\mu\nu\rho\sigma}$, and examine their behaviour in the region of the classical ring singularity at $r=0$ and $\theta=\frac{\pi}{2}$. Analytic expressions for these two quantities are given in Appendix \ref{singularities}.

We begin with the Ricci scalar where classically we have $R=0$. Now, we substitute \eqref{Gofr} and find that a divergence at $r=0$, $\theta=\frac{\pi}{2}$ as 
\begin{equation}
R\sim\frac{1}{\epsilon}
\end{equation}
in the limit $\epsilon\to0$, where $\epsilon=\sqrt{r^2+a^2\cos\theta}$. We observe that this corresponds to a ring singularity. For the Kretschmann invariant we have
\begin{equation}
K\sim\frac{1}{\epsilon^2}
\end{equation}
for $\epsilon\to0$. We also observe that the singularity is significantly softened compared to the classical case where $R\sim\epsilon^{1-d}$ and the Kretschmann invariant diverges as $K\sim\epsilon^{2-2d}$.

In order to study the effects of different matching conditions between momentum and position scales on the behaviour of the singularities we approximate  the running of the gravitational coupling near the origin by
\begin{equation}\label{sigma}
M\,G(r)=\mu^{\sigma}\,r^{\sigma+d-3},
\end{equation}
following \cite{kevin}, where $\mu$ is a parameter fixed by the renormalisation group with dimensions of mass. Note that $\sigma=1$ is the matching used throughout this paper, whereas $\sigma = 1-d$ reduces \eq{sigma} to the classical theory.  Then, by substituting this into our expressions in Appendix \ref{singularities} we find that the Ricci scalar diverges as
\begin{equation}
R\sim\frac{1}{\epsilon^{2-\sigma}}
\end{equation}
and the Kretschmann invariant as
\begin{equation}
K\sim\frac{1}{\epsilon^{4-2\sigma}}.
\end{equation}
The above results indicate that quantum corrections have the property to significantly soften the black hole singularities. Incidentally, as shown in \cite{kevin} for non-rotating black holes, this softening is sufficient to make the space-time geodesically complete.  By using slightly different matchings we observe that the singularities could be removed completely, and we point out that the space-time is regular when $\sigma\geq2$. Similar results have  been obtained for the non-rotating case  \cite{reuter}, \cite{kevin}. We conclude that the addition of rotation does not substantially alter the singularity structure.

\subsection{Mass and angular momentum}

In stationary, asymptotically flat space-times we can use the Killing vectors associated with time translations and rotations to define the total mass and the total angular momentum of the space-time. This is done by associating a conserved charge to each Killing vector through the Komar integrals \cite{komar}   
\begin{equation}
Q_{\xi}(\partial\Sigma)=c \oint_{\partial \Sigma} \nabla^{\mu}\xi^{\nu}d\Sigma_{\mu\nu}\label{komarcharge}
\end{equation}
where $\xi$ is the Killing vector, $\Sigma$ is a spacelike hypersurface, $\partial \Sigma$ its boundary, $d\Sigma_{\mu\nu}$ is the surface element of $\partial \Sigma$ and $c$ is a constant. In order to get the total mass of the space-time, the boundary $\partial \Sigma$, a two-sphere, is taken at infinity or at any exterior vacuum region. These expressions were generalized in \cite{Hawking2} for a space-time containing a black hole and has been given in terms of a boundary integral at the horizon and a hypersurface integral at the region between the horizon and $\partial \Sigma_{\infty}$ at infinity. After fixing the constant $c$, we arrive at the following expression for the mass of a stationary asymptotically flat space-time in $d$ dimensions \cite{MyersPerry}
\begin{equation}
M=-\frac{1}{16\pi G_N}\frac{d-2}{d-3}\left[ 2\int_\Sigma R^{\mu}_{\nu}k^{\nu}d\Sigma_{\mu} + \oint_{\mathcal{H}}\nabla^{\mu} k^{\nu}d\Sigma_{\mu\nu}\right] \label{komarmass}
\end{equation}
where the integral over $\Sigma$ is performed from the horizon until $\partial \Sigma_{\infty}$ and $d\Sigma_{\mu}$ is the surface element on $\Sigma$. We consider the first integral as it gives the contribution to the total mass of the matter outside the event horizon and the second integral as the mass of the black hole. Using Einstein's equations one can express the first integral in terms of the energy momentum tensor as $\int_{\Sigma} \left( T^{\mu}_{\nu}k^{\nu}-\frac{1}{2}Tk^{\mu}\right) d\Sigma_{\mu}$. Similarly, the angular momentum of a stationary and asymptotically flat space-time is defined as
\begin{equation}
J=-\frac{1}{16\pi G_N}\left[ 2\int_\Sigma R^{\mu}_{\nu}m^{\nu}d\Sigma_{\mu} + \oint_{\mathcal{H}}\nabla^{\mu} m^{\nu}d\Sigma_{\mu\nu}\right] \label{komarangmom}.
\end{equation}
We recall that classical Myers-Perry black holes are vacuum solutions of Einstein's equations. Then, it is straightforward to find that the Komar integrals performed at the horizon return the physical mass and angular momentum of the black hole. As we saw in the previous section, our space-time features an effective energy momentum tensor and we would like to know how this affects the mass and the angular momentum of the black holes. In what follows we are going to denote the boundary integrals at the horizon by 
\begin{equation}
\begin{split}
M_{\mathcal{H}}=&-\frac{1}{16\pi G_N}\frac{d-2}{d-3}\oint_{\mathcal{H}}\nabla^{\mu} k^{\nu}d\Sigma_{\mu\nu}\\
J_{\mathcal{H}}=&-\frac{1}{16\pi G_N}\oint_{\mathcal{H}}\nabla^{\mu} m^{\nu}d\Sigma_{\mu\nu}.\label{MJKomar}
\end{split}
\end{equation}
At first, we would like to verify that the total mass and angular momentum of the space-time remain the same. To do this we perform the integral \eqref{komarcharge} at $\partial\Sigma_{\infty}$ for the two Killing vectors and we find that they indeed return $Q_k(\partial\Sigma_{\infty})=M_{\textnormal{phys}}$ and $Q_m(\partial\Sigma_{\infty})=\frac{2}{d-2}\,a\, M_{\textnormal{phys}}=J$.

It is interesting to see the modifications to the mass and angular momentum of the black hole due to quantum corrections. Performing the boundary integral for the timelike Killing vector, we find the following expression for the mass of a black hole 
\begin{equation}
M_{\mathcal{H}}=M_{\textnormal{phys}}\frac{G(r_+)}{G_N}\left[1+\frac{\eta(r_+)}{d-3}\cdot F\right],\label{komarmassH}
\end{equation}
where $F=\leftexp{2}{F}_{1}\left(1,1;\frac{d-1}{2};\frac{a^2}{r_+^2+a^2}\right)$ is the Gaussian hypergeometric function. By substituting $d=4$ in \eqref{komarmassH} we confirm the Komar mass of the four dimensional black holes as found previously in \cite{tuiran} for a specific form of the RG running. 

We note some interesting properties of this expression. Since all its parameters are positive, the hypergeometric function in \eqref{komarmassH} will be positive. Thus, for a theory where the gravitational coupling becomes weaker (so $\eta(r)$ is negative) the mass of the RG-improved black hole will be lower than the classical mass parameter $M_{\textnormal{phys}}$. 

We are now interested in how $M_{\mathcal{H}}$ varies when we change the parameters $a$ and $g_*$, which control the kinematical and quantum correcitons, respectively. We find that as either of them grow the Komar mass gets smaller. Thus, the black holes will reach their minimum masses at criticality. Moreover, it is easily shown that  the Komar mass reaches an absolute minimum value $M_{\mathcal{H}}=0$ when the black holes are in one of the two following critical configurations. First, for every dimension, for a critical non-rotating black hole we have $M_{\mathcal{H}}=0$, since the term inside the square brackets becomes $1+\eta(r_{\textnormal{c}})/(d-3)$, which vanishes. The second case where  the horizon mass vanishes is when we have a critical ultra-spinning black hole. In this case we have $\frac{r_{\textnormal{c}}}{a}\to0$ and again we conclude from \eqref{komarmassH} that $M_{\mathcal{H}}\to0$. 

Similarly, we now compute the Komar integral for the Killing vector $m$ to find the angular momentum of the black holes. This is given by the expression
\begin{equation}
J_{\mathcal{H}}=J\frac{G(r_+)}{G_N}\left[1+\frac{1}{2}\eta(r_+)\left(\frac{r_+^2}{a^2}+1\right)\left(F-1 \right)\right].
\end{equation}
Again, the limit $d=4$ confirms the result given in \cite{tuiran} for the Komar angular momentum of the RG-improved Kerr solution for a specific choice of the RG running. Inspection of the above formula shows that the angular momentum of the horizon is always less than the classical value $J$. However, in contrast to the Komar mass, this formula can turn negative for some values of the parameters, which implies that the effective rotation of the horizon is in the opposite direction.

\subsection{Remarks on the laws of black hole mechanics}

Having reviewed the basic thermodynamical properties of the quantum corrected space-times, we are able to discuss their implications to the laws of black hole mechanics. Here, we are going to briefly comment on possible deviations from these classical laws.

First, we want to examine the validity of the integral formula 
\begin{equation}
M_{\mathcal{H}} -2\Omega_HJ_{\mathcal{H}}=\frac{\kappa}{4\pi}\mathcal{A}_H. \label{IntegralFormula}
\end{equation}
This is the analogue of Smarr's formula  \cite{smarr} for a stationary axisymmetric space-time (not necessary in vacuum) which contains a black hole. Then, the values of mass, angular momentum and area at the horizon are related through the relation \eqref{IntegralFormula}. It follows directly from the derivation of \cite{Hawking2} that the integral formula is a concequence only of the properties of the Killing vectors and of  the constancy of surface gravity on the horizon. Thus, we expect that in the case of quantum corrections parametrized by $G(r)$ this will still hold true. Indeed, we can also verify the validity of the integral formula by using the expressions of $M_{\mathcal{H}}$ and $J_{\mathcal{H}}$ obtained in the previous section.

The zeroth law of black hole thermodynamics states that the surface gravity of a stationary black hole is constant over the event horizon. It is easily seen, directly from the expression for the surface gravity
\begin{equation}
\kappa=\frac{1}{2\,r_+}\left(\frac{2\,r^2_+}{r_+^2+a^2}+d-5+\eta(r_+)\right),
\end{equation}
that the zeroth law holds true in the RG-improved case as well. It is interesting to note, that the proof given in \cite{Hawking2} relies on the dominant energy condition. However, there are alternative proofs of the zeroth law \cite{carter,Wald}, which, instead, rely on the existence of a bifurcate Killing horizon. Moreover, the proofs also work in the opposite direction, implying that if the surface gravity is constant then there exists a bifurcate Killing horizon.  

In the previous two cases we saw that classical relations hold when we consider asymptotically safe black holes. However, this is not in general true for the first law of black hole mechanics. This is the differential law relating variations of the mass, the angular momentum and the area of the black hole
\begin{equation}
dM= \frac{\kappa}{4\pi}d\mathcal{A}_H+\Omega_HdJ.\label{firstlaw}
\end{equation}
In classical Einstein gravity this relation, was first used to identify the entropy of the black hole with the area of the horizon, $S=\mathcal{A}_H/(4\pi G_N)$. Corrections to this simple form of the entropy are well known to exist in theories of modified gravity or when quantum corrections are considered and various techniques have been developed for its calculation \cite{WaldEntropy,Thermo,Jacobson:1993vj}.  It has also been argued that the first law of black hole thermodynamics persists under RG corrections provided the matching condition accounts for both quantum and kinematical effects \cite{kevin2}.

In what concerns us here, we note from the original derivation \cite{Hawking2} that  extra contributions arise provided that Einstein's equations imply an effective EM tensor. These additional contributions take the form 
\begin{equation}
dM= \frac{\kappa}{4\pi}d\mathcal{A}_H+\Omega_HdJ_H + d\int{ T^{\mu}_{\nu}k^{\nu}d\Sigma_{\mu}}.\label{firstlawQ}
\end{equation}
The term involving the energy-momentum tensor gives a contribution from the angular momentum outside the horizon and also contributions from the energy density and the pressures of the effective matter. It is evident that in general the first law need not to hold true in its classical form given by \eqref{firstlaw}. In order to define the entropy we should in principle include the additional contributions following the general procedure highlighted in \cite{Gibbons:1976ue,Entropy1}. This is left for future work.

Recently, by studying the four-dimensional Kerr black hole using RG corrections analogous to those implemented here \cite{tuiran}, the authors suggested that the standard laws of black hole thermodynamics cannot hold true. This finding is based on the observation that no integrating factor can be found to make the 
one-form state function $S(M,J)$ exact.
Subsequently, in \cite{kevin2} it was shown that using a thermodynamically-motivated class of matching conditions  between the momentum and position space, the first law of black hole mechanics holds in its classical form given by \eqref{firstlaw}.

Finally, further investigation requires also the second law of black hole mechanics \cite{hawking3} which states that the area of the event horizon of a black hole does not decrease with time
\begin{equation}
d\mathcal{A}_H\geq0.
\end{equation}
The proof of the second law in general relativity relies on the requirement that the energy momentum tensor of the space-time obeys the dominant energy condition. As we have seen, the dominant energy condition does no longer hold true for the entire space-time as soon as RG corrections of $G(r)$ are taken into account, including those studied here.

\section{Discussion}\label{Conclusions}

We have studied quantum-gravitational corrections to rotating black holes, based on an RG improved version of Myers-Perry space-times.  The horizon structure crucially depends on whether gravity strengthens or weakens towards shorter distances. Provided that gravity weakens according to the asymptotic safety conjecture, we find a smallest achievable black hole mass $M_c(a)$ for given rotation $a=J/M$.
The critical black hole mass is a dynamically-generated short-distance effect,  dictated by the underlying UV fixed point $g_*$, and with a mild dependence on angular momentum.
 The physics interpretation of this feature relates to the fact that asymptotic safety predicts a weakening of gravity at shortest scales. Then, for small black holes, gravity is no longer strong enough to generate an event horizon, or to counteract rotation. The resulting smallest, critical black holes are cold, and imply that Myers-Perry black holes, rotating or not, display an upper bound on temperature. Furthermore, this pattern is largely independent on the specific form of the matching condition employed to link the RG running of couplings to the black hole geometry, and, thus, a stable prediction of our theory.

Quantitatively, in the semi-classical limit where the mass of the black hole is much larger than the fluctuation-induced critical mass $M_c/M\ll 1$, quantum corrections to the horizon radii, temperature and specific heat remain perturbatively small. For smaller black holes, corrections become of order one, temperature reaches a maximum, and the specific heat changes sign.  The ring singularity inside the black hole persists, albeit in a substantially softened version.  In \cite{kevin}, it was observed that this softening is sufficient to make RG-improved Schwarzschild space-times geodesically complete, suggesting that a mild singularity may persist even for quantum black holes.  Also, the effective energy momentum tensor does not satisfy the weak and dominant energy condition in the entire space-time, suggesting that the laws of black hole thermodynamics are modified as well.  Interestingly, our findings  agree with complementary studies exploiting the relations imposed by black hole thermodynamics \cite{kevin2}, thus adding to the overall consistency of the picture.

It remains to discuss 
the interplay between kinematical effects induced by rotation, and fluctuation-induced effects encoded by the running of Newton's coupling. 
Classically, rotation lifts a degeneracy of the Schwarzschild black hole in four dimensions by generating a Cauchy horizon in addition to the event horizon. In five and more dimensions, however, the Cauchy horizon disappears, and in six and more dimensions classical Myers-Perry black holes can even accommodate arbitrarily high angular momentum. Quantum-mechanically, fluctuations lift a degeneracy  of Schwarzschild black holes in any dimension \cite{reuter,kevin}. In addition, the Cauchy horizon then also encodes the conformal scaling of the underlying gravitational fixed point \cite{kevin2}. 
  As has been shown in detail here, fluctuations also lift the degeneracy of Myers-Perry black holes by generating both an event horizon and a Cauchy horizon in {\it any} dimension. 
We therefore conclude that quantum-gravitational corrections dominate over kinematical ones such as those induced by rotation. 

\acknowledgements

We thank Kevin Falls for discussions. This work is supported by the Science Technology and Facilities Council (STFC) under grant number [ST/J000477/1], and by the A.S.~Onassis Public Benefit Foundation under grant number [F-ZG066/2010-2011].

\appendix

\section{Horizons}\label{Appendix}

In this Appendix we are going to show that under the assumptions discussed in section \ref{generalremarks} the space-time has always either two horizons (an event and a Cauchy horizon), one extreme, or no horizons at all depending on the values of its parameters.

First we want to define what are the assumptions for $G(r)$ that we are going to use and to extend them, so to cover the maximum possible variety of runnings. For this we review the relation which defines the horizons for non-rotating black holes. This reads
\begin{equation}
\left.f(r)\right|_{a=0}=1-\frac{M\,G(r)}{r^{d-3}}=0
\end{equation}
and its first derivative (which is the first derivative of the gravitational potential) is given by
\begin{equation}
V'(r)=M\, G(r)\, r^{2-d}\left[ d-3+\eta(r)\right].
\end{equation}
In order for non-rotating black holes to have the usual behavior (either two, one critical or no horizons) we have to demand two things. First, that the limit $r\to0$ of the lapse function is positive. This implies that
\begin{equation}
\lim_{r\to0}V(r)=\lim_{r\to0}\left(-\frac{M\,G(r)}{r^{d-3}}\right)>-1.\label{assump1}
\end{equation}
If this hold true then the limits of $f(r)$ as $r\to0$ and as $r\to\infty$ are both positive. Consequently, if $f(r)$ has only one minimum, the space-time has two horizons if this minimum is negative, one degenerate horizon if this minimum is $0$ and no horizons if it is positive.

In order to achieve this behavior we need that $V'(r)$ changes from negative to positive values only once. For this to be true it is enough to postulate that the anomalous dimension for gravity $\eta(r)$ is a monotonically increasing function of $r$ and that it satisfies
\begin{equation}
\lim_{r\to0}\eta(r)<3-d, \qquad \qquad \lim_{r\to\infty}\eta(r)=0. \label{assump2}
\end{equation}
Then, it is evident that if \eqref{assump1} and \eqref{assump2} hold, the space-time for non-rotating black holes has the desired behavior. 

In the case of asymptotic safety, the above two assumptions can be reduced to a single assumption for the matching of momentum and position scales. Follwing \cite{kevin}, we make the identification
\begin{equation}
k(r)\sim\frac{\xi}{r^{\gamma}}
\end{equation}
where $\xi$ has a non-trivial mass dimension if $\gamma\neq 1$, and use the running for $G(k)$ implied by asymptotic safety \eqref{Gk}, then the two assumptions reduce to the single condition for the parameter $\gamma$
\begin{equation}
\gamma>\frac{d-3}{d-2}\,. \label{assumptiongamma}
\end{equation}
We note, that the matching we are commonly using for our calculations is $k(r)=1/r$, meaning $\gamma=1$, which clearly satisfies the condition \eqref{assumptiongamma}.

Now we are going to see that the above assumptions are enough to ensure that the space-time of rotating black holes still has the same horizon structure in four and five dimensions. For six or higher dimensions we will see that we have to demand more details about the behavior of $G(r)$. Now, the relation which gives the horizons is
\begin{equation}
f(r)=1+\frac{a^2}{r^2}-\frac{M\,G(r)}{r^{d-3}}. 
\end{equation}
The limits of this function for $r\to0$ and $r\to\infty$ are again both positive under the condition \eqref{assump1}. Again, we need to show that this function has only one minimum and so that its first derivative changes from negative to positive values only once. So, we turn our attention to 
\begin{equation}
r^3\,f'(r)=-2a^2+M\, G(r)\, r^{5-d}\left[ d-3+\eta(r)\right]. \label{fprime}
\end{equation}
In what follows we will need to find out also about the behaviour of the second term in \eqref{fprime}, so we define the function $U(r)=M\, G(r)\, r^{5-d}\left[ d-3+\eta(r)\right]$ and we write down its first derivative $U'(r)=-M\, G(r)\, r^{4-d}\left[\left(d-5+\eta(r)\right)\left(d-3+\eta(r)\right)-r\,\eta'(r)\right]$. Moreover, we define $r_1$ as the value of $r$ for which the anomalous dimension becomes $\eta(r_1)=3-d$ and $r_2$ when we have $\eta(r_2)=5-d$. Now, we need to distinguish between $d=4$, $d=5$ and $d\geq6$ in order to find the behavior of $f'(r)$. We start with the four dimensional case.

\paragraph{Four dimensions.}
Using both assumptions \eqref{assump1} and \eqref{assump2} we can see that the limit of $f'(r)$ as $r\to0$ is negative and the limit of $f'(r)$ as $r\to \infty$ is positive. However, we need to know that $f'(r)$ changes sign only once. For the region of $r$ between $0\leq r \leq r_1$, $f'(r)$ is always negative. So, in order to change sign only once we have to assure that for $r>r_1$, $U(r)$ is always a growing function of $r$. This is easily seen by looking at its first derivative $U'(r)$, since for $r>r_1$ then $\eta(r)>-1$ and $\eta'(r)>0$ always.

\paragraph{Five dimensions.}
In five dimensions the limit $r\to0$ of $f'(r)$ is still negative, but the limit $r\to\infty$ has the sign of $M\, G_N-a^2$. This being positive, is just the condition for the existence of roots in the classical case. In the case where $M\, G_N-a^2>0$, the limit of $f'(r)$ when $r\to\infty$ is positive and is easily checked again from $U'(r)$ that for $r>r_1$, $U(r)$ is growing and so that the space-time has the expected behavior. If on the other hand $M\, G_N-a^2<0$ then $f'(r)$ is always negative and there are no horizons.

\paragraph{Six and more dimensions.}
Now the limit $r\to0$ of $f'(r)$ is still negative, but the limit $r\to\infty$ is negative too. In order to have only one minimum for $f(r)$ we need that $f'(r)$ has the most two roots. This is satisfied if for $r>r_1$, $U(r)$ grows until a maximum value and then decreases to $0$. For this we have to check the sign of the expression inside square brackets in $U'(r)$. For $r_1<r<r_2$ then $U(r)$ is positive implying that $U(r)$ grows. Now, in order for rotating black holes in $d\geq6$ dimensions to have the desired horizon structure we have to impose the condition that for $r>r_2$ the function $U'(r)$ has only one root. This condition is far stronger in terms of $\eta(r)$ than those that we have assumed so far, but in the case of asymptotic safety and for a running given by \eqref{Gk} with a matching that obeys $\gamma>\frac{d-3}{d-2} $ it is easily checked that this condition holds.

For completeness we state a derivation of the fact that the function of $G(k)$ given by \eqref{Gk} using the linear matching $k\sim1/r$ gives the usual horizon structure with two, one critical or no horizons. In what follows we use dimensionless variables and we start looking for roots of the function $\tilde\Delta(x)$ (with $G_N$ substituted by $G(x)$) 
\begin{equation}
\tilde\Delta(x)=A+x^2-\frac{x^3}{x^{d-2}+\Omega}\label{corectedDelta}
\end{equation}
It is obvious, that now the limits $x\to0$ and $x\to\infty$, (in contrast to the classical case for $d\geq6$) do not guarantee the existence of a horizon in any dimensionality, since they always return $\tilde\Delta=A$ and $\tilde\Delta=\infty$ respectively. The next step is to look at the first derivative of $\tilde\Delta$ with respect to $x$, which is written
\begin{equation}
\begin{split}
\tilde\Delta'(x)=&\frac{x}{(x^{d-2}+\Omega)^2}\cdot\left[2x^{2(d-2)}+(d-5)x^{d-1}\right.\\
&\left.+4\Omega x^{d-2}-3\Omega x+2\Omega^2\right]\label{deltaprime}
\end{split}
\end{equation}
At first sight this doesn't look very helpful. In order to have horizons it is necessary (but not sufficient) that $\Delta'(x)$ becomes negative for some $x$. The limits $x\to0$ and $x\to\infty$ of $\tilde\Delta'(x)$ give $0$ and $\infty$ respectively, providing as with no information about the roots of $\tilde\Delta'(x)$. 

Now we note, that the roots of $\tilde\Delta'(x)$ are the roots of the expression in square brackets  in \eqref{deltaprime}, which we call $N(x)$, and moreover that the sign of $\tilde\Delta'(x)$ is that of $N(x)$. A careful study of $N(x)$ and its derivatives shows that, for $d\geq5$, it starts from a positive value ($2\Omega^2$), it decreases until some value $N(x_1)$ and then increases up to infinity. Whether or not $N(x_1)$ (and therefore $\tilde\Delta'(x_1)$) is negative, depends (for each dimensionality $d$) only on the value of $\Omega$. If $\tilde\Delta'(x)$ becomes negative, then it has two roots $x_2$ and $x_0$, with $x_2<x_1< x_0$.    

With this information we have what we need to determine the behavior of $\tilde\Delta(x)$. It begins from the positive value $A$ and it starts increasing. If $\tilde\Delta'(x)$ remains positive for every $x$, then $\tilde\Delta(x)$ continues to increase and there are no horizons. If $\tilde\Delta'(x)$ becomes negative at some interval $(x_2,x_0)$, then $\tilde\Delta(x)$ decreases until a value $\tilde\Delta(x_0)$ and then increases to infinity. When $\tilde\Delta(x_0)<0$ the space-time has two horizons. If $\tilde\Delta(x_0)=0$ the space-time has one degenerate horizon, while if $\tilde\Delta(x_0)>0$ there are no horizons.

\section{Energy-momentum tensor}\label{Appendix2}
The coefficients $U^{\mu}_{\nu}$  in equation \eqref{UandV} are given by
\begin{equation}
\begin{split}
&U^{t}_{t}=\frac{Mr^{4-d}}{8\Sigma^3}\,\left[(9-2d)a^4+2(8-3d)a^2r^2-4(d-2)r^4-2(d-4)a^2(a^2+r^2)\cos(2\theta)-a^4\cos(4\theta)\right]\\
&U^{r}_{r}=-\frac{Mr^{4-d}}{2\Sigma^2}\,\left[(d-2)r^2+(d-4)a^2\cos^2\theta\right]\\
&U^{\theta}_{\theta}=-\frac{Mr^{4-d}}{\Sigma^2} \,a^2\cos^2\theta\\
&U^{\phi}_{\phi}=\frac{Mr^{4-d}}{8\Sigma^3}\,a^2\left[1/2(d-12)a^2+2(d-4)r^2-2(2a^2+dr^2)\cos(2\theta)-1/2(d-4)a^2\cos(4\theta)\right]\\
&U^{t}_{\phi}=-\frac{Mr^{4-d}}{2\Sigma^3}\,a(a^2+r^2)\sin^2\theta\,\left[(d-2)r^2+(d-6)a^2\cos^2\theta\right]\\
&U^{\phi}_t=\frac{Mr^{4-d}}{2\Sigma^3}\,a\left[(d-2)r^2+(d-6)a^2\cos^2\theta\right]\\
&U^{i}_{i}=0. \label{UMUnu}
\end{split}
\end{equation}
Similarly, for the coefficients $V^{\mu}_{\nu}$ we get the following expressions
\begin{equation}
\begin{split}
&V^{t}_{t}=\frac{Mr^{4-d}}{2\Sigma^2}\,a^2r\sin^2{\theta}\\
&V^{r}_{r}=0\\
&V^{\theta}_{\theta}=-\frac{Mr^{4-d}}{2\Sigma}\,r\\
&V^{\phi}_{\phi}=-\frac{Mr^{4-d}}{2\Sigma^2}\,r(a^2+r^2)\\
&V^{t}_{\phi}=\frac{Mr^{4-d}}{2\Sigma^2}\,a \,r(a^2+r^2)\sin^2\theta\\
&V^{\phi}_t=-\frac{Mr^{4-d}}{2\Sigma^2}\,a\, r \\
&V^{i}_{i}=-\frac{Mr^{4-d}}{2\Sigma}\,r.\label{VMUnu}
\end{split}
\end{equation}
To check the energy conditions we have to diagonalise the EM tensor. The resulting diagonal $T^{\mu\textnormal{(eff)}}_{\nu\textnormal{(diag)}}$ is 
\begin{equation}
T^{\mu\textnormal{(eff)}}_{\nu\textnormal{(diag)}}=\frac{Mr^{4-d}}{8\Sigma^3}\textnormal{diag}\left(T^0_0,T^{r}_{r},T^{\theta}_{\theta},T^3_3, T^4_4, \ldots,T^{d-1}_{d-1} \right)
\end{equation}
whith  $T^0_0$ and $T^3_3$ being the diagonalised components given by 
\begin{eqnarray}
T^0_0&=\frac{1}{2}\left(T^{t}_{t}+T^{\phi}_{\phi}-\sqrt{\left(T^{t}_{t}-T^{\phi}_{\phi}\right)^2+4T^{t}_{\phi}T^{\phi}_t}\right)\\
T^3_3&=\frac{1}{2}\left(T^{t}_{t}+T^{\phi}_{\phi}+\sqrt{\left(T^{t}_{t}-T^{\phi}_{\phi}\right)^2+4T^{t}_{\phi}T^{\phi}_t}\right).
\end{eqnarray}
After substituting the expressions for the energy momentum components from equations \eqref{UMUnu} and \eqref{VMUnu} we get for $T^0_0$ and $T^3_3$
\begin{eqnarray}
T^0_0&=\left\{
\begin{array}{rl}
-\frac{Mr^{4-d}}{2\Sigma^2}\left[(d-2)r^2+(d-4)a^2\cos^2\theta\right]G'(r) &  \,\,\,\,  \text{if } Z(r)>0,\\
-\frac{Mr^{4-d}}{2\Sigma^2}\left[2a^2\cos^2\theta\,G'(r)+r\Sigma G''(r)\right] & \,\,\,\,  \text{if } Z(r) < 0.
\end{array} \right.\\
T^3_3&=\left\{
\begin{array}{rl}
-\frac{Mr^{4-d}}{2\Sigma^2}\left[2a^2\cos^2\theta\,G'(r)+r\Sigma G''(r)\right] &  \,\,\,\,  \text{if } Z(r) > 0,\\
-\frac{Mr^{4-d}}{2\Sigma^2}\left[(d-2)r^2+(d-4)a^2\cos^2\theta\right]G'(r) & \,\,\,\,  \text{if } Z(r)<0.
\end{array} \right.
\end{eqnarray}
Here $Z(r)$ is an expression which has \emph{always one} positive real root for a matching of the form \eqref{Gofr} and is given by
\begin{equation}
Z(r)=2\left[(d-2)r^2+(d-6)a^2\cos^2\theta\right]G'(r)-2\,r\,\Sigma\, G''(r).
\end{equation}
It is  essential to know at every point of the space-time which is the timelike component of the diagonalised energy-momentum tensor which corresponds to the energy density. For this, we take the eigenvectors $v^{\mu}_1$ and $v^{\mu}_2$ used to diagonalise $T^{\mu\textnormal{(eff)}}_{\nu}$ and we compute $g_{\mu\nu}v^{\mu}_1 v^{\nu}_1$ and $g_{\mu\nu}v^{\mu}_2 v^{\nu}_2$. We find that when $r$ is between the two horizons, $T^r_r$ is the timelike component. When $r$ is not between the two horizons and $Z(r)>0$ the timelike component is $T^3_3$, while when $r$ is not between the horizons and $Z(r)<0$ the timelike component is $T^0_0$. 
In any of these three cases the diagonal EM-tensor takes the form
\begin{equation}
T^{\mu\textnormal{(eff)}}_{\nu\textnormal{(diag)}}=\textnormal{diag}\left(-\rho,p_r,p_{\perp},p_{\perp},p_4, \ldots,p_{d-1}\right),
\end{equation}
where the timelike component is always given by
\begin{equation}
\rho=\frac{M\,r^{4-d}}{2\Sigma^2}\left[(d-2)r^2+(d-4)a^2\cos^2\theta\right]G'(r)\label{Density}
\end{equation}
and all the other components correspond to spacelike coordinates, with
\begin{equation}
p_{\perp}=-\frac{M\,r^{4-d}}{2\Sigma^2}\left[2a^2\cos^2\theta \,G'(r)+r\Sigma G''(r)\right].
\end{equation}
and 
\begin{equation}
p_r=-\rho=T^r_r, \qquad \qquad p_i=T^i_i.\label{Pressure}
\end{equation}

\section{Kretschmann invariant}\label{singularities}

Here, we compute two of the curvature invariants, the Ricci scalar and the Kretschmann invariant and we examine the fate of the classical ring singularity at $r=0, \theta=\pi/2$.
The Ricci scalar in our case takes the form
\begin{equation}
R=M\,r^{4-d}\cdot\frac{r\,G''(r)+2\,G'(r)}{r^2+a^2\cos^2\theta},
\end{equation}
while the Kretschmann invariant $K=R_{\mu\nu\rho\sigma}R^{\mu\nu\rho\sigma}$ is given by the following formula
\begin{equation}
\begin{split}
K=&\frac{M^2\,r^{6-2d}}{\left(r^2+a^2\cos^2\theta\right)^6}\cdot\left[K_1\, G(r)^2+K_2\,G(r)\,G'(r)+K_3\,G(r)\,G''(r)\right.\\
&\left.+K_4\, G'(r)^2+K_5\, G'(r)\, G''(r)+ K_6\, G''(r)^2\right]\label{Kretschmann}
\end{split}
\end{equation}
with the coefficients $K_1$, $K_2$, $K_3$, $K_4$, $K_5$ and $K_6$ given by
\begin{equation}
\begin{split}
K_1=&(d-3) (d-2)^2 (d-1)\, r^8+4(d-1) \left(-15+(-d+5) (d-4) \,d\right)\, r^6\,a^2 \cos^2\theta+\\
& 6 \left(22+(d-6) \left(7+(d-6) \,d\right)\,d\right) \,r^4\,a^4 \cos^4\theta+4 (d-5) \left(3+(d-5) (d-4) \,d\right)\, r^2\, a^6\cos^6\theta+\\
& (d-5) (d-4)^2 (d-3)\,a^8 \cos^8\theta \\
&\\
K_2=&-4\, r \,(r^2 + a^2\cos^2\theta)\cdot\left[ (d-3) (d-2)^2 r^6+\left(-48+d \left(64+3 (d-9)\, d\right)\right)\, r^4\, a^2\cos^2\theta+\right.\\
& \left. \left(-68+d \left(104+3 (d-11)\, d\right)\right)\, r^2\,a^4\cos^4\theta+(d-5) (d-4)^2\, a^6\cos^6\theta\right]\\
&\\
K_3=&2\, r^2 (r^2 + a^2\cos^2\theta)^2\cdot\left[(d-3) (d-2)\, r^4+2 (9+(d-7) d)\, r^2 \,a^2\cos^2\theta+\right.\\
&\left.(d-5) (d-4)a^4\cos^4\theta\right] \\
&\\
K_4=&2\, r^2 (r^2 + a^2\cos^2\theta)\cdot\left[(16+d (2 d-11))\, r^4+2 (23+d (2 d-15)) \,r^2\,a^2\cos^2\theta)+\right.\\
&\left.(46+d\, (2 d-19))\, a^4\cos^4\theta)\right]\\
&\\
K_5=&16\, r \,(r^2 + a^2\cos^2\theta)\cdot\left[ (d-3)\, r^2+(d-5)\,a^2\cos^2\theta\right]\\
&\\
K_6=& 32\,r^3\, (r^2 + a^2\cos^2\theta)^3
\end{split}
\end{equation}
If we approximate the running of the gravitational coupling near the origin as
$G(r)=\mu^{\sigma}\,r^{\sigma+d-3}$,
we find the following expression in terms of $\sigma$
\begin{equation}
\begin{split}
K=&\frac{16\,M^2\,\mu^2\,r^{2\sigma}}{\Sigma^6}\left[384 \, r^8-192 r^6 (\sigma+4 ) \Sigma +8\, r^4 \left(51+2 d+38 \sigma +6 \sigma ^2\right)\Sigma^2\right.\\
&- 4\, r^2 (\sigma +2) (3+2 d+2 \sigma  (5+\sigma )) \Sigma^3\\
&\left.+\left(12+2 d^2+2 d \,(\sigma-1 ) (\sigma +5)+\sigma  (-20+\sigma  (\sigma +1) (\sigma +5))\right)\Sigma^4\right]
\end{split}
\end{equation}
and for the Ricci scalar
\begin{equation}
R=\frac{M\, \mu\, r^{\sigma } (d+\sigma -3) (d+\sigma -2)}{r^2+a^2\cos^2\theta}
\end{equation}

\bibliography{paper}

\begin{thebibliography}{63}
\expandafter\ifx\csname natexlab\endcsname\relax\def\natexlab#1{#1}\fi
\expandafter\ifx\csname bibnamefont\endcsname\relax
  \def\bibnamefont#1{#1}\fi
\expandafter\ifx\csname bibfnamefont\endcsname\relax
  \def\bibfnamefont#1{#1}\fi
\expandafter\ifx\csname citenamefont\endcsname\relax
  \def\citenamefont#1{#1}\fi
\expandafter\ifx\csname url\endcsname\relax
  \def\url#1{\texttt{#1}}\fi
\expandafter\ifx\csname urlprefix\endcsname\relax\def\urlprefix{URL }\fi
\providecommand{\bibinfo}[2]{#2}
\providecommand{\eprint}[2][]{\url{#2}}

\bibitem[{\citenamefont{Schwarzschild}(1916)}]{Schwarzschild:1916uq}
\bibinfo{author}{\bibfnamefont{K.}~\bibnamefont{Schwarzschild}},
  \bibinfo{journal}{Sitzungsber.Preuss.Akad.Wiss.Berlin (Math.Phys.)} pp.
  \bibinfo{pages}{189--196} (\bibinfo{year}{1916}).

\bibitem[{\citenamefont{Kerr}(1963)}]{Kerr:1963ud}
\bibinfo{author}{\bibfnamefont{R.~P.} \bibnamefont{Kerr}},
  \bibinfo{journal}{Phys.Rev.Lett.} \textbf{\bibinfo{volume}{11}},
  \bibinfo{pages}{237} (\bibinfo{year}{1963}).

\bibitem[{\citenamefont{Tangherlini}(1963)}]{Tangherlini}
\bibinfo{author}{\bibfnamefont{F.~R.} \bibnamefont{Tangherlini}},
  \bibinfo{journal}{Nuovo Cim.} \textbf{\bibinfo{volume}{27}},
  \bibinfo{pages}{636} (\bibinfo{year}{1963}).

\bibitem[{\citenamefont{Myers and Perry}(1986)}]{MyersPerry}
\bibinfo{author}{\bibfnamefont{R.~C.} \bibnamefont{Myers}} \bibnamefont{and}
  \bibinfo{author}{\bibfnamefont{M.~J.} \bibnamefont{Perry}},
  \bibinfo{journal}{Ann. Phys.} \textbf{\bibinfo{volume}{172}},
  \bibinfo{pages}{304} (\bibinfo{year}{1986}).

\bibitem[{\citenamefont{Emparan and Myers}(2003)}]{instabus}
\bibinfo{author}{\bibfnamefont{R.}~\bibnamefont{Emparan}} \bibnamefont{and}
  \bibinfo{author}{\bibfnamefont{R.~C.} \bibnamefont{Myers}},
  \bibinfo{journal}{JHEP} \textbf{\bibinfo{volume}{09}} (\bibinfo{year}{2003}),
  \eprint{hep-th/0308056}.

\bibitem[{\citenamefont{Dias et~al.}(2010{\natexlab{a}})\citenamefont{Dias,
  Figueras, Monteiro, Reall, and Santos}}]{Dias:2010eu}
\bibinfo{author}{\bibfnamefont{O.~J.} \bibnamefont{Dias}},
  \bibinfo{author}{\bibfnamefont{P.}~\bibnamefont{Figueras}},
  \bibinfo{author}{\bibfnamefont{R.}~\bibnamefont{Monteiro}},
  \bibinfo{author}{\bibfnamefont{H.~S.} \bibnamefont{Reall}}, \bibnamefont{and}
  \bibinfo{author}{\bibfnamefont{J.~E.} \bibnamefont{Santos}},
  \bibinfo{journal}{JHEP} \textbf{\bibinfo{volume}{1005}}, \bibinfo{pages}{076}
  (\bibinfo{year}{2010}{\natexlab{a}}), \eprint{hep-th/1001.4527}.

\bibitem[{\citenamefont{Dias et~al.}(2009)\citenamefont{Dias, Figueras,
  Monteiro, Santos, and Emparan}}]{Dias:2009iu}
\bibinfo{author}{\bibfnamefont{O.~J.} \bibnamefont{Dias}},
  \bibinfo{author}{\bibfnamefont{P.}~\bibnamefont{Figueras}},
  \bibinfo{author}{\bibfnamefont{R.}~\bibnamefont{Monteiro}},
  \bibinfo{author}{\bibfnamefont{J.~E.} \bibnamefont{Santos}},
  \bibnamefont{and} \bibinfo{author}{\bibfnamefont{R.}~\bibnamefont{Emparan}},
  \bibinfo{journal}{Phys.Rev.} \textbf{\bibinfo{volume}{D80}},
  \bibinfo{pages}{111701} (\bibinfo{year}{2009}), \eprint{hep-th/0907.2248}.

\bibitem[{\citenamefont{Gubser and Mitra}(2001)}]{Gubser:2000mm}
\bibinfo{author}{\bibfnamefont{S.~S.} \bibnamefont{Gubser}} \bibnamefont{and}
  \bibinfo{author}{\bibfnamefont{I.}~\bibnamefont{Mitra}},
  \bibinfo{journal}{JHEP} \textbf{\bibinfo{volume}{0108}}, \bibinfo{pages}{018}
  (\bibinfo{year}{2001}), \eprint{hep-th/0011127}.

\bibitem[{\citenamefont{Dias et~al.}(2010{\natexlab{b}})\citenamefont{Dias,
  Figueras, Monteiro, and Santos}}]{Dias:2010maa}
\bibinfo{author}{\bibfnamefont{O.~J.} \bibnamefont{Dias}},
  \bibinfo{author}{\bibfnamefont{P.}~\bibnamefont{Figueras}},
  \bibinfo{author}{\bibfnamefont{R.}~\bibnamefont{Monteiro}}, \bibnamefont{and}
  \bibinfo{author}{\bibfnamefont{J.~E.} \bibnamefont{Santos}},
  \bibinfo{journal}{Phys.Rev.} \textbf{\bibinfo{volume}{D82}},
  \bibinfo{pages}{104025} (\bibinfo{year}{2010}{\natexlab{b}}),
  \eprint{hep-th/1006.1904}.

\bibitem[{\citenamefont{Monteiro et~al.}(2009)\citenamefont{Monteiro, Perry,
  and Santos}}]{Monteiro:2009tc}
\bibinfo{author}{\bibfnamefont{R.}~\bibnamefont{Monteiro}},
  \bibinfo{author}{\bibfnamefont{M.~J.} \bibnamefont{Perry}}, \bibnamefont{and}
  \bibinfo{author}{\bibfnamefont{J.~E.} \bibnamefont{Santos}},
  \bibinfo{journal}{Phys.Rev.} \textbf{\bibinfo{volume}{D80}},
  \bibinfo{pages}{024041} (\bibinfo{year}{2009}), \eprint{gr-qc/0903.3256}.

\bibitem[{\citenamefont{Emparan et~al.}(2007)\citenamefont{Emparan, Harmark,
  Niarchos, Obers, and Rodriguez}}]{phasestructure}
\bibinfo{author}{\bibfnamefont{R.}~\bibnamefont{Emparan}},
  \bibinfo{author}{\bibfnamefont{T.}~\bibnamefont{Harmark}},
  \bibinfo{author}{\bibfnamefont{V.}~\bibnamefont{Niarchos}},
  \bibinfo{author}{\bibfnamefont{N.~A.} \bibnamefont{Obers}}, \bibnamefont{and}
  \bibinfo{author}{\bibfnamefont{M.~J.} \bibnamefont{Rodriguez}},
  \bibinfo{journal}{JHEP} \textbf{\bibinfo{volume}{10}} (\bibinfo{year}{2007}),
  \eprint{hep-th/0708.2181}.

\bibitem[{\citenamefont{Emparan and Reall}(2002)}]{blackrings}
\bibinfo{author}{\bibfnamefont{R.}~\bibnamefont{Emparan}} \bibnamefont{and}
  \bibinfo{author}{\bibfnamefont{H.~S.} \bibnamefont{Reall}},
  \bibinfo{journal}{Phys. Rev. Lett.} \textbf{\bibinfo{volume}{88}}
  (\bibinfo{year}{2002}), \eprint{hep-th/0110260}.

\bibitem[{\citenamefont{'t~Hooft and Veltman}(1974)}]{tHooft:1974bx}
\bibinfo{author}{\bibfnamefont{G.}~\bibnamefont{'t~Hooft}} \bibnamefont{and}
  \bibinfo{author}{\bibfnamefont{M.}~\bibnamefont{Veltman}},
  \bibinfo{journal}{Annales Poincare Phys.Theor.}
  \textbf{\bibinfo{volume}{A20}}, \bibinfo{pages}{69} (\bibinfo{year}{1974}).

\bibitem[{\citenamefont{Weinberg}(1979)}]{Weinberg:1980gg}
\bibinfo{author}{\bibfnamefont{S.}~\bibnamefont{Weinberg}}
  (\bibinfo{year}{1979}), \bibinfo{note}{in: General Relativity: An Einstein
  centenary survey, Eds. S.W. Hawking and W. Israel, Cambridge University
  Press, p. 790}.

\bibitem[{\citenamefont{Reuter}(1998)}]{Reuter:1996cp}
\bibinfo{author}{\bibfnamefont{M.}~\bibnamefont{Reuter}},
  \bibinfo{journal}{Phys.Rev.} \textbf{\bibinfo{volume}{D57}},
  \bibinfo{pages}{971} (\bibinfo{year}{1998}), \eprint{hep-th/9605030}.

\bibitem[{\citenamefont{Souma}(1999)}]{Souma:1999at}
\bibinfo{author}{\bibfnamefont{W.}~\bibnamefont{Souma}},
  \bibinfo{journal}{Prog.Theor.Phys.} \textbf{\bibinfo{volume}{102}},
  \bibinfo{pages}{181} (\bibinfo{year}{1999}), \eprint{hep-th/9907027}.

\bibitem[{\citenamefont{Falls et~al.}(2013)\citenamefont{Falls, Litim,
  Nikolakopoulos, and Rahmede}}]{Falls:2013bv}
\bibinfo{author}{\bibfnamefont{K.}~\bibnamefont{Falls}},
  \bibinfo{author}{\bibfnamefont{D.}~\bibnamefont{Litim}},
  \bibinfo{author}{\bibfnamefont{K.}~\bibnamefont{Nikolakopoulos}},
  \bibnamefont{and} \bibinfo{author}{\bibfnamefont{C.}~\bibnamefont{Rahmede}}
  (\bibinfo{year}{2013}), \eprint{hep-th/1301.4191}.

\bibitem[{\citenamefont{Litim}(2004)}]{Litim:2003vp}
\bibinfo{author}{\bibfnamefont{D.~F.} \bibnamefont{Litim}},
  \bibinfo{journal}{Phys.Rev.Lett.} \textbf{\bibinfo{volume}{92}},
  \bibinfo{pages}{201301} (\bibinfo{year}{2004}), \eprint{hep-th/0312114}.

\bibitem[{\citenamefont{Litim}(2006)}]{Litim:2006dx}
\bibinfo{author}{\bibfnamefont{D.~F.} \bibnamefont{Litim}},
  \bibinfo{journal}{AIP Conf.Proc.} \textbf{\bibinfo{volume}{841}},
  \bibinfo{pages}{322} (\bibinfo{year}{2006}), \eprint{hep-th/0606044}.

\bibitem[{\citenamefont{Fischer and
  Litim}(2006{\natexlab{a}})}]{Fischer:2006fz}
\bibinfo{author}{\bibfnamefont{P.}~\bibnamefont{Fischer}} \bibnamefont{and}
  \bibinfo{author}{\bibfnamefont{D.~F.} \bibnamefont{Litim}},
  \bibinfo{journal}{Phys.Lett.} \textbf{\bibinfo{volume}{B638}},
  \bibinfo{pages}{497} (\bibinfo{year}{2006}{\natexlab{a}}),
  \eprint{hep-th/0602203}.

\bibitem[{\citenamefont{Lauscher and Reuter}(2002)}]{Lauscher:2002sq}
\bibinfo{author}{\bibfnamefont{O.}~\bibnamefont{Lauscher}} \bibnamefont{and}
  \bibinfo{author}{\bibfnamefont{M.}~\bibnamefont{Reuter}},
  \bibinfo{journal}{Phys.Rev.} \textbf{\bibinfo{volume}{D66}},
  \bibinfo{pages}{025026} (\bibinfo{year}{2002}), \eprint{hep-th/0205062}.

\bibitem[{\citenamefont{Codello et~al.}(2008)\citenamefont{Codello, Percacci,
  and Rahmede}}]{Codello:2007bd}
\bibinfo{author}{\bibfnamefont{A.}~\bibnamefont{Codello}},
  \bibinfo{author}{\bibfnamefont{R.}~\bibnamefont{Percacci}}, \bibnamefont{and}
  \bibinfo{author}{\bibfnamefont{C.}~\bibnamefont{Rahmede}},
  \bibinfo{journal}{Int.J.Mod.Phys.} \textbf{\bibinfo{volume}{A23}},
  \bibinfo{pages}{143} (\bibinfo{year}{2008}), \eprint{hep-th/0705.1769}.

\bibitem[{\citenamefont{Machado and Saueressig}(2008)}]{Machado:2007ea}
\bibinfo{author}{\bibfnamefont{P.~F.} \bibnamefont{Machado}} \bibnamefont{and}
  \bibinfo{author}{\bibfnamefont{F.}~\bibnamefont{Saueressig}},
  \bibinfo{journal}{Phys.Rev.} \textbf{\bibinfo{volume}{D77}},
  \bibinfo{pages}{124045} (\bibinfo{year}{2008}), \eprint{hep-th/0712.0445}.

\bibitem[{\citenamefont{Benedetti et~al.}(2009)\citenamefont{Benedetti,
  Machado, and Saueressig}}]{Benedetti:2009rx}
\bibinfo{author}{\bibfnamefont{D.}~\bibnamefont{Benedetti}},
  \bibinfo{author}{\bibfnamefont{P.~F.} \bibnamefont{Machado}},
  \bibnamefont{and}
  \bibinfo{author}{\bibfnamefont{F.}~\bibnamefont{Saueressig}},
  \bibinfo{journal}{Mod.Phys.Lett.} \textbf{\bibinfo{volume}{A24}},
  \bibinfo{pages}{2233} (\bibinfo{year}{2009}), \eprint{hep-th/0901.2984}.

\bibitem[{\citenamefont{Percacci and Perini}(2003)}]{Percacci:2002ie}
\bibinfo{author}{\bibfnamefont{R.}~\bibnamefont{Percacci}} \bibnamefont{and}
  \bibinfo{author}{\bibfnamefont{D.}~\bibnamefont{Perini}},
  \bibinfo{journal}{Phys.Rev.} \textbf{\bibinfo{volume}{D67}},
  \bibinfo{pages}{081503} (\bibinfo{year}{2003}), \eprint{hep-th/0207033}.

\bibitem[{\citenamefont{Folkerts et~al.}(2012)\citenamefont{Folkerts, Litim,
  and Pawlowski}}]{Folkerts:2011jz}
\bibinfo{author}{\bibfnamefont{S.}~\bibnamefont{Folkerts}},
  \bibinfo{author}{\bibfnamefont{D.~F.} \bibnamefont{Litim}}, \bibnamefont{and}
  \bibinfo{author}{\bibfnamefont{J.~M.} \bibnamefont{Pawlowski}},
  \bibinfo{journal}{Phys.Lett.} \textbf{\bibinfo{volume}{B709}},
  \bibinfo{pages}{234} (\bibinfo{year}{2012}), \eprint{hep-th/1101.5552}.

\bibitem[{\citenamefont{Harst and Reuter}(2011)}]{Harst:2011zx}
\bibinfo{author}{\bibfnamefont{U.}~\bibnamefont{Harst}} \bibnamefont{and}
  \bibinfo{author}{\bibfnamefont{M.}~\bibnamefont{Reuter}},
  \bibinfo{journal}{JHEP} \textbf{\bibinfo{volume}{1105}}, \bibinfo{pages}{119}
  (\bibinfo{year}{2011}), \eprint{hep-th/1101.6007}.

\bibitem[{\citenamefont{Zanusso et~al.}(2010)\citenamefont{Zanusso, Zambelli,
  Vacca, and Percacci}}]{Zanusso:2009bs}
\bibinfo{author}{\bibfnamefont{O.}~\bibnamefont{Zanusso}},
  \bibinfo{author}{\bibfnamefont{L.}~\bibnamefont{Zambelli}},
  \bibinfo{author}{\bibfnamefont{G.}~\bibnamefont{Vacca}}, \bibnamefont{and}
  \bibinfo{author}{\bibfnamefont{R.}~\bibnamefont{Percacci}},
  \bibinfo{journal}{Phys.Lett.} \textbf{\bibinfo{volume}{B689}},
  \bibinfo{pages}{90} (\bibinfo{year}{2010}), \eprint{hep-th/0904.0938}.

\bibitem[{\citenamefont{Litim}(2011)}]{Litim:2011cp}
\bibinfo{author}{\bibfnamefont{D.~F.} \bibnamefont{Litim}},
  \bibinfo{journal}{Phil.Trans.Roy.Soc.Lond.} \textbf{\bibinfo{volume}{A369}},
  \bibinfo{pages}{2759} (\bibinfo{year}{2011}), \eprint{1102.4624}.

\bibitem[{\citenamefont{Litim}(2008)}]{Litim:2008tt}
\bibinfo{author}{\bibfnamefont{D.~F.} \bibnamefont{Litim}}
  (\bibinfo{year}{2008}), \eprint{0810.3675}.

\bibitem[{\citenamefont{Fischer and
  Litim}(2006{\natexlab{b}})}]{Fischer:2006at}
\bibinfo{author}{\bibfnamefont{P.}~\bibnamefont{Fischer}} \bibnamefont{and}
  \bibinfo{author}{\bibfnamefont{D.~F.} \bibnamefont{Litim}},
  \bibinfo{journal}{AIP Conf.Proc.} \textbf{\bibinfo{volume}{861}},
  \bibinfo{pages}{336} (\bibinfo{year}{2006}{\natexlab{b}}),
  \eprint{hep-th/0606135}.

\bibitem[{\citenamefont{Niedermaier and Reuter}(2006)}]{Niedermaier:2006wt}
\bibinfo{author}{\bibfnamefont{M.}~\bibnamefont{Niedermaier}} \bibnamefont{and}
  \bibinfo{author}{\bibfnamefont{M.}~\bibnamefont{Reuter}},
  \bibinfo{journal}{Living Rev.Rel.} \textbf{\bibinfo{volume}{9}},
  \bibinfo{pages}{5} (\bibinfo{year}{2006}).

\bibitem[{\citenamefont{Codello et~al.}(2009)\citenamefont{Codello, Percacci,
  and Rahmede}}]{Codello:2008vh}
\bibinfo{author}{\bibfnamefont{A.}~\bibnamefont{Codello}},
  \bibinfo{author}{\bibfnamefont{R.}~\bibnamefont{Percacci}}, \bibnamefont{and}
  \bibinfo{author}{\bibfnamefont{C.}~\bibnamefont{Rahmede}},
  \bibinfo{journal}{Annals Phys.} \textbf{\bibinfo{volume}{324}},
  \bibinfo{pages}{414} (\bibinfo{year}{2009}), \eprint{hep-th/0805.2909}.

\bibitem[{\citenamefont{Bonanno and Reuter}(2000)}]{reuter}
\bibinfo{author}{\bibfnamefont{A.}~\bibnamefont{Bonanno}} \bibnamefont{and}
  \bibinfo{author}{\bibfnamefont{M.}~\bibnamefont{Reuter}},
  \bibinfo{journal}{Phys. Rev.} \textbf{\bibinfo{volume}{D62}}
  (\bibinfo{year}{2000}), \eprint{hep-th/0002196}.

\bibitem[{\citenamefont{Falls et~al.}(2012)\citenamefont{Falls, Litim, and
  Raghuraman}}]{kevin}
\bibinfo{author}{\bibfnamefont{K.}~\bibnamefont{Falls}},
  \bibinfo{author}{\bibfnamefont{D.~F.} \bibnamefont{Litim}}, \bibnamefont{and}
  \bibinfo{author}{\bibfnamefont{A.}~\bibnamefont{Raghuraman}},
  \bibinfo{journal}{Int.J.Mod.Phys.} \textbf{\bibinfo{volume}{A27}},
  \bibinfo{pages}{1250019} (\bibinfo{year}{2012}), \eprint{hep-th/1002.0260}.

\bibitem[{\citenamefont{Reuter and Tuiran}(2006)}]{Reuter:2006rg}
\bibinfo{author}{\bibfnamefont{M.}~\bibnamefont{Reuter}} \bibnamefont{and}
  \bibinfo{author}{\bibfnamefont{E.}~\bibnamefont{Tuiran}}, pp.
  \bibinfo{pages}{2608--2610} (\bibinfo{year}{2006}), \eprint{hep-th/0612037}.

\bibitem[{\citenamefont{Reuter and Tuiran}(2011)}]{tuiran}
\bibinfo{author}{\bibfnamefont{M.}~\bibnamefont{Reuter}} \bibnamefont{and}
  \bibinfo{author}{\bibfnamefont{E.}~\bibnamefont{Tuiran}},
  \bibinfo{journal}{Phys.Rev. D} \textbf{\bibinfo{volume}{83}}
  (\bibinfo{year}{2011}), \eprint{hep-th/1009.3528}.

\bibitem[{\citenamefont{Cai and Easson}(2010)}]{Cai:2010zh}
\bibinfo{author}{\bibfnamefont{Y.-F.} \bibnamefont{Cai}} \bibnamefont{and}
  \bibinfo{author}{\bibfnamefont{D.~A.} \bibnamefont{Easson}},
  \bibinfo{journal}{JCAP} \textbf{\bibinfo{volume}{1009}}, \bibinfo{pages}{002}
  (\bibinfo{year}{2010}), \eprint{hep-th/1007.1317}.

\bibitem[{\citenamefont{Becker and Reuter}(2012)}]{Becker:2012js}
\bibinfo{author}{\bibfnamefont{D.}~\bibnamefont{Becker}} \bibnamefont{and}
  \bibinfo{author}{\bibfnamefont{M.}~\bibnamefont{Reuter}},
  \bibinfo{journal}{JHEP} \textbf{\bibinfo{volume}{1207}}, \bibinfo{pages}{172}
  (\bibinfo{year}{2012}), \eprint{hep-th/1205.3583}.

\bibitem[{\citenamefont{Falls and Litim}(2012)}]{kevin2}
\bibinfo{author}{\bibfnamefont{K.}~\bibnamefont{Falls}} \bibnamefont{and}
  \bibinfo{author}{\bibfnamefont{D.~F.} \bibnamefont{Litim}}
  (\bibinfo{year}{2012}), \eprint{gr-qc/1212.1821}.

\bibitem[{\citenamefont{Bennett}(2009)}]{Bennett}
\bibinfo{author}{\bibfnamefont{S.}~\bibnamefont{Bennett}},
  \bibinfo{journal}{MSc Thesis, U. Sussex}  (\bibinfo{year}{2009}).

\bibitem[{\citenamefont{Koch and Saueressig}(2013)}]{Koch:2013owa}
\bibinfo{author}{\bibfnamefont{B.}~\bibnamefont{Koch}} \bibnamefont{and}
  \bibinfo{author}{\bibfnamefont{F.}~\bibnamefont{Saueressig}}
  (\bibinfo{year}{2013}), \eprint{1306.1546}.

\bibitem[{\citenamefont{Emparan and Reall}(2008)}]{Emparan}
\bibinfo{author}{\bibfnamefont{R.}~\bibnamefont{Emparan}} \bibnamefont{and}
  \bibinfo{author}{\bibfnamefont{H.~S.} \bibnamefont{Reall}},
  \bibinfo{journal}{Living Rev. Rel.} \textbf{\bibinfo{volume}{11}},
  \bibinfo{pages}{6} (\bibinfo{year}{2008}), \eprint{hep-th/0801.3471}.

\bibitem[{\citenamefont{Litim}(2001)}]{Litim:2001up}
\bibinfo{author}{\bibfnamefont{D.~F.} \bibnamefont{Litim}},
  \bibinfo{journal}{Phys.Rev.} \textbf{\bibinfo{volume}{D64}},
  \bibinfo{pages}{105007} (\bibinfo{year}{2001}), \eprint{hep-th/0103195}.

\bibitem[{\citenamefont{Gerwick et~al.}(2011)\citenamefont{Gerwick, Litim, and
  Plehn}}]{Gerwick:2011jw}
\bibinfo{author}{\bibfnamefont{E.}~\bibnamefont{Gerwick}},
  \bibinfo{author}{\bibfnamefont{D.}~\bibnamefont{Litim}}, \bibnamefont{and}
  \bibinfo{author}{\bibfnamefont{T.}~\bibnamefont{Plehn}},
  \bibinfo{journal}{Phys.Rev.} \textbf{\bibinfo{volume}{D83}},
  \bibinfo{pages}{084048} (\bibinfo{year}{2011}), \eprint{hep-th/1101.5548}.

\bibitem[{\citenamefont{Nicolini et~al.}(2006)\citenamefont{Nicolini,
  Smailagic, and Spallucci}}]{Nicolini:2005vd}
\bibinfo{author}{\bibfnamefont{P.}~\bibnamefont{Nicolini}},
  \bibinfo{author}{\bibfnamefont{A.}~\bibnamefont{Smailagic}},
  \bibnamefont{and}
  \bibinfo{author}{\bibfnamefont{E.}~\bibnamefont{Spallucci}},
  \bibinfo{journal}{Phys.Lett.} \textbf{\bibinfo{volume}{B632}},
  \bibinfo{pages}{547} (\bibinfo{year}{2006}), \eprint{gr-qc/0510112}.

\bibitem[{\citenamefont{Modesto and Nicolini}(2010)}]{Modesto:2010rv}
\bibinfo{author}{\bibfnamefont{L.}~\bibnamefont{Modesto}} \bibnamefont{and}
  \bibinfo{author}{\bibfnamefont{P.}~\bibnamefont{Nicolini}},
  \bibinfo{journal}{Phys.Rev.} \textbf{\bibinfo{volume}{D82}},
  \bibinfo{pages}{104035} (\bibinfo{year}{2010}), \eprint{gr-qc/1005.5605}.

\bibitem[{\citenamefont{Christodoulou}(1970)}]{Christodoulou:1970wf}
\bibinfo{author}{\bibfnamefont{D.}~\bibnamefont{Christodoulou}},
  \bibinfo{journal}{Phys.Rev.Lett.} \textbf{\bibinfo{volume}{25}},
  \bibinfo{pages}{1596} (\bibinfo{year}{1970}).

\bibitem[{\citenamefont{Christodoulou and
  Ruffini}(1971)}]{Christodoulou:1972kt}
\bibinfo{author}{\bibfnamefont{D.}~\bibnamefont{Christodoulou}}
  \bibnamefont{and} \bibinfo{author}{\bibfnamefont{R.}~\bibnamefont{Ruffini}},
  \bibinfo{journal}{Phys.Rev.} \textbf{\bibinfo{volume}{D4}},
  \bibinfo{pages}{3552} (\bibinfo{year}{1971}).

\bibitem[{\citenamefont{Hawking}(1975)}]{Hawking}
\bibinfo{author}{\bibfnamefont{S.~W.} \bibnamefont{Hawking}},
  \bibinfo{journal}{Commun. Math. Phys.} \textbf{\bibinfo{volume}{43}},
  \bibinfo{pages}{199} (\bibinfo{year}{1975}).

\bibitem[{\citenamefont{Gibbons and Hawking}(1977)}]{Gibbons:1976ue}
\bibinfo{author}{\bibfnamefont{G.~W.} \bibnamefont{Gibbons}} \bibnamefont{and}
  \bibinfo{author}{\bibfnamefont{S.~W.} \bibnamefont{Hawking}},
  \bibinfo{journal}{Phys. Rev.} \textbf{\bibinfo{volume}{D15}},
  \bibinfo{pages}{2752} (\bibinfo{year}{1977}).

\bibitem[{\citenamefont{Birrell and Davies}(1982)}]{Birrell:1982ix}
\bibinfo{author}{\bibfnamefont{N.}~\bibnamefont{Birrell}} \bibnamefont{and}
  \bibinfo{author}{\bibfnamefont{P.}~\bibnamefont{Davies}},
  \emph{\bibinfo{title}{{Quantum fields in curved space}}}
  (\bibinfo{year}{1982}).

\bibitem[{\citenamefont{Jacobson et~al.}(1994)\citenamefont{Jacobson, Kang, and
  Myers}}]{Jacobson:1993vj}
\bibinfo{author}{\bibfnamefont{T.}~\bibnamefont{Jacobson}},
  \bibinfo{author}{\bibfnamefont{G.}~\bibnamefont{Kang}}, \bibnamefont{and}
  \bibinfo{author}{\bibfnamefont{R.~C.} \bibnamefont{Myers}},
  \bibinfo{journal}{Phys.Rev.} \textbf{\bibinfo{volume}{D49}},
  \bibinfo{pages}{6587} (\bibinfo{year}{1994}), \eprint{gr-qc/9312023}.

\bibitem[{\citenamefont{Hawking and Ellis}(1973)}]{Hawking:1973uf}
\bibinfo{author}{\bibfnamefont{S.}~\bibnamefont{Hawking}} \bibnamefont{and}
  \bibinfo{author}{\bibfnamefont{G.}~\bibnamefont{Ellis}},
  \emph{\bibinfo{title}{{The Large scale structure of space-time}}}
  (\bibinfo{year}{1973}).

\bibitem[{\citenamefont{Komar}(1959)}]{komar}
\bibinfo{author}{\bibfnamefont{A.}~\bibnamefont{Komar}},
  \bibinfo{journal}{Phys. Rev.} \textbf{\bibinfo{volume}{113}},
  \bibinfo{pages}{934} (\bibinfo{year}{1959}).

\bibitem[{\citenamefont{Bardeen et~al.}(1973)\citenamefont{Bardeen, Carter, and
  Hawking}}]{Hawking2}
\bibinfo{author}{\bibfnamefont{J.~M.} \bibnamefont{Bardeen}},
  \bibinfo{author}{\bibfnamefont{B.}~\bibnamefont{Carter}}, \bibnamefont{and}
  \bibinfo{author}{\bibfnamefont{S.~W.} \bibnamefont{Hawking}},
  \bibinfo{journal}{Commun. Math. Phys.} \textbf{\bibinfo{volume}{31}},
  \bibinfo{pages}{161} (\bibinfo{year}{1973}).

\bibitem[{\citenamefont{Smarr}(1973)}]{smarr}
\bibinfo{author}{\bibfnamefont{L.}~\bibnamefont{Smarr}},
  \bibinfo{journal}{Phys. Rev. Lett.} \textbf{\bibinfo{volume}{30}},
  \bibinfo{pages}{71} (\bibinfo{year}{1973}).

\bibitem[{\citenamefont{Carter}(1973)}]{carter}
\bibinfo{author}{\bibfnamefont{B.}~\bibnamefont{Carter}},
  \bibinfo{journal}{Gordon and Breach Science Publishers, New
  York--London--Paris} pp. \bibinfo{pages}{125--214} (\bibinfo{year}{1973}).

\bibitem[{\citenamefont{Racz and Wald}(1992)}]{Wald}
\bibinfo{author}{\bibfnamefont{I.}~\bibnamefont{Racz}} \bibnamefont{and}
  \bibinfo{author}{\bibfnamefont{R.~M.} \bibnamefont{Wald}},
  \bibinfo{journal}{Class. Quant. Grav.} \textbf{\bibinfo{volume}{9}},
  \bibinfo{pages}{2643} (\bibinfo{year}{1992}).

\bibitem[{\citenamefont{Wald}(1993)}]{WaldEntropy}
\bibinfo{author}{\bibfnamefont{R.~M.} \bibnamefont{Wald}},
  \bibinfo{journal}{Phys.Rev.} \textbf{\bibinfo{volume}{D48}},
  \bibinfo{pages}{3427} (\bibinfo{year}{1993}).

\bibitem[{\citenamefont{Hawking and Page}(1983)}]{Thermo}
\bibinfo{author}{\bibfnamefont{S.~W.} \bibnamefont{Hawking}} \bibnamefont{and}
  \bibinfo{author}{\bibfnamefont{D.~N.} \bibnamefont{Page}},
  \bibinfo{journal}{Commun. Math. Phys.} \textbf{\bibinfo{volume}{87}},
  \bibinfo{pages}{577} (\bibinfo{year}{1983}).

\bibitem[{\citenamefont{Visser}(1993)}]{Entropy1}
\bibinfo{author}{\bibfnamefont{M.}~\bibnamefont{Visser}},
  \bibinfo{journal}{Phys.Rev.} \textbf{\bibinfo{volume}{D48}},
  \bibinfo{pages}{583} (\bibinfo{year}{1993}).

\bibitem[{\citenamefont{Hawking}(1972)}]{hawking3}
\bibinfo{author}{\bibfnamefont{S.~W.} \bibnamefont{Hawking}},
  \bibinfo{journal}{Commun. Math. Phys.} \textbf{\bibinfo{volume}{26}}
  (\bibinfo{year}{1972}).

\end{thebibliography}

\end{document}